\documentclass[sigconf,nonacm]{acmart}
\setcopyright{none}

\newcommand{\ignore}[1]{}
\usepackage{xspace}
\usepackage{wrapfig}
\usepackage{subfig}
\usepackage{pifont}
\usepackage{xcolor}

\usepackage{csquotes}
\ignore{
\def\csq@setfrcodes{%
    \ifnum\sfcode`\A=\@m
    \else
        \csq@setazcodes
    \fi
    \sfcode`\,=1003
    \sfcode`\;=1004
    \sfcode`\:=1005
    \sfcode`\.=1006
    \sfcode`\!=1007
    \sfcode`\?=1008
}
}
\makeatletter
\DeclareRobustCommand{\change}{%
  \@bsphack
  \leavevmode
  \color{red}%
  \@esphack
}
\DeclareRobustCommand{\stopchange}{%
  \@bsphack
  \normalcolor
  \@esphack
}
\makeatother

\newcommand{\name}{\hmm{T}urnpike\xspace}
\newcommand\hmm[1]{\ifnum\ifhmode\spacefactor\else2000\fi>1004 \uppercase{#1}\else#1\fi}

\newcommand{\HJ}[1]{{\color{red}\bfseries [[HJ: #1]]}}
\newcommand{\HJdel}[1]{{\color{red}\sout{#1}}}
\newcommand{\revision}[1]{#1}

\makeatletter
\newcommand*{\rom}[1]{\expandafter\@slowromancap\romannumeral #1@}
\makeatother

\AtBeginDocument{%
  \providecommand\BibTeX{{%
      \normalfont B\kern-0.5em{\scshape i\kern-0.25em b}\kern-0.8em\TeX}}}

\begin{document}

\title{Lightweight Soft Error Resilience for In-Order Cores}
\author{Jianping Zeng}
  \affiliation{
  \institution{Purdue University}
  \country{USA}
}
\email{zeng207@purdue.edu}
\author{Hongjune Kim}
  \affiliation{
  \institution{Seoul National University}
  \country{Korea}
}
\email{hongjune@aces.snu.ac.kr}
\author{Jaejin Lee}
  \affiliation{
  \institution{Seoul National University}
  \country{Korea}
}
\email{jaejin@snu.ac.kr}
\author{Changhee Jung}
  \affiliation{
  \institution{Purdue University}
  \country{USA}
}
\email{chjung@purdue.edu}

\begin{abstract}
Acoustic-sensor-based soft error resilience is particularly promising, since it
can verify the absence of soft errors and eliminate silent data corruptions
at a low hardware cost.  However, the state-of-the-art work incurs a
significant performance overhead for in-order cores due to frequent structural/data hazards during the verification. To address the problem, this paper
presents \name, a compiler/architecture co-design scheme that can achieve
lightweight yet guaranteed soft error resilience for in-order cores. The key
idea is that many of the data computed in the core can bypass the soft error
verification without compromising the resilience.  Along with simple
microarchitectural support for realizing the idea, \name leverages compiler
optimizations to further reduce the performance overhead.  Experimental results
with 36 benchmarks demonstrate that \name only incurs a 0-\revision{14\%} run-time overhead
on average while the state-of-the-art incurs a 29-84\% overhead when the worst-case latency of the sensor based error detection is 10-50 cycles.

\end{abstract}

\maketitle

\section{Introduction}
\label{sec:intro}
Soft error resilience is becoming more important than ever. With technology
scaling, circuits are likely to be more sensitive to radiation-induced soft errors; they
are mostly caused by energetic particles (e.g., cosmic rays) and 
 alpha particles from packaging materials~\cite{jang11,mukherjee05,UpasaniVG12,UpasaniVG13,UpasaniVG14,UpasaniVG15,DeBardeleben14}.
Soft errors may lead to a system crash or even worse silent data
corruptions (SDC) that are not caught by the error detection logic but end up
with incorrect outputs. Due to the high availability requirement of
embedded systems, soft error resilience has been one of the most important
design considerations.

Among existing soft error resilience schemes, acoustic-sensor-based
detection~\cite{UpasaniVG12,UpasaniVG13,UpasaniVG14,UpasaniVG15,upasani2016soft,upasani2016case,naveed2020aster,chen2020compiler,liu2016low,liu2015clover,liu2017compiler,liu2018compiler}
is particularly promising. To the best of our knowledge, it is the only way to prevent
SDC---that is a long-awaited open problem---at a low hardware cost. Since acoustic sensors
perceive the sound wave of particle strikes, which is always generated as a physical
phenomenon, no resulting soft error is missed. As such, the sensor-based detection can
achieve SDC freedom; unlike other schemes, it does not even require any 
microarchitecture replication. 
Moreover, sensors occupy only a very small die size area.
For example, 300 sensors are enough to achieve 30 cycles of the
worst-case detection latency (WCDL) for a 2GHz out-of-order core, and they
only cause $\sim$1\% area overhead~\cite{UpasaniVG12,UpasaniVG13,UpasaniVG14,UpasaniVG15,upasani2016soft,upasani2016case}.

With that in mind, Liu~\emph{et~al}~\cite{liu2016low} show how their
solution Turnstile can leverage acoustic sensors for core-level error
containment with little architecture change for soft error
verification/recovery. The rationale for verifying the absence of soft errors
is that since each error is to be detected within WCDL after its occurrence,
execution prior to a given time $T$ will be verified to be error-free
at a time $T$+WCDL, if no error is detected during the WCDL.  

In light of this, Turnstile verifies every data being stored to memory, ensuring that it
has not been affected by soft errors before its write-back. Although a re-order buffer (ROB)
retires a store instructions, the data is not written back to memory but held in
a store buffer until it turns out to be verified waiting for WCDL. 
%
%
For register verification, Turnstile reformulates it based on the aforementioned memory
verification by inserting stores to checkpoint updated live-out registers and
holding them in the store buffer.
The upshot is that register write-backs are never delayed for verification,
which would otherwise slow down the pipeline execution. Since
stores are rarely on the critical path in out-of-order cores, Turnstile can
offer lightweight soft error resilience at $\approx$8\% performance overhead
on average for SPEC2006/MediaBench/SPLASH2 benchmarks.

Compared to out-of-order (OoO) cores, however, there has been less attention received
to enhance the reliability of in-order cores in a low-cost manner---though they
are widely deployed in embedded systems to control the physical world. For
example, while in-order cores are used for adaptive cruise control, precrash
safety alarm, and motion planning systems due to the simple hardware 
and the time predictability demand excluding complex OoO execution~\cite{hahn2019design,
venkataramani2020time,mische2010enhance,schoeberl2011towards,schoeberl2015t,durrieu2014predictable},
they still rely on expensive dual/triple modular redundancy (DMR/TMR) to deal with soft errors~\cite{ces2020arm,
iturbe2019arm,instruments2018advanced,berntorp2019motion,reality2018arm,
takahashi20164,hernandez2015timely,mehra2015adaptive}. Nonetheless, 
mission-critical embedded systems should pursue power-efficiency as they are often battery operated, e.g., portable military devices, wearables, and drones, preventing the use of DMR/TMR. Apart from that, DMR/TMR could suffer the size and weight issues that are particularly critical for tiny aerial systems such as spying drones~\cite{nassi2021sok, nassi2019drones,clothier2015risk} and bionic birds~\cite{bionicbird,zhu2015study}.

With the increasing demand for lightweight soft error resilience for in-order cores,
one might want to leverage Turnstile on top of in-order cores. Unfortunately, 
naively adapting Turnstile to in-order cores causes a significant performance overhead,
i.e., 29\%-84\% for 10-50 cycles of WCDL. The main reason is that the in-order pipeline
stalls for the structural/data hazards of stores due to the inability to schedule other
independent instructions.  Since the store buffer of in-order cores is very small (4 
entries) unlike that of out-of-order cores (40 or more entries), it often becomes full
during the verification, in which case the pipeline stalls on the next store due to
the structural hazard until some of the buffered stores are verified and flushed to L1 cache.
%
Similarly, for the execution of a checkpoint, i.e., essentially a store
instruction to save a register value, the in-order pipeline may stall waiting
for the value to be available. This data hazard happens a lot leading to significant performance degradation, because Turnstile inserts the checkpoint right after the register-update instruction, e.g., a delinquent load.

To address the problems, this paper presents \name, a compiler/architecture
co-design scheme that can achieve a lightweight yet guaranteed soft error
resilience for in-order cores. \name leverages 3 key insights to minimize the pipeline stalls with lowering the store buffer pressure.
First, during code generation, it is possible to decrease the number of stores to be verified.
\name's compiler optimizations 
remove unnecessary checkpoint stores, e.g., those whose value can be
reconstructed from other checkpointed values at the recovery time~\cite{pennyPLDI20,liu2016compiler},
without compromising the recoverability.
We also propose 2 novel compiler
optimizations to suppress the generation of stores: loop induction variable merging for reducing live registers being checkpointed in a loop, and store-aware register allocation for less register-spilling stores.

Second, the compiler can reduce the execution delay of unremoved checkpoint stores with the help of 
instruction scheduling for resolving the checkpoint data hazard.
That is, \name attempts to separate the live register-update instructions from their dependent checkpoint stores by filling the gap with other independent instructions.
This gives the in-order core an illusion that it can hide the execution delay of the checkpoint
as in out-of-order execution.

Third, many of the remaining stores can be safely released to cache without
waiting for verification, no matter if they are regular stores or checkpoint stores. For example, some value being stored is never used for the recovery of a soft error---even if it corrupts the value. To take advantage
of this insight, we introduce simple hardware support that can (1) conduct the
safety check for the fast (early) release of a given store and, if possible,
(2) let it go through the fast path\footnote{The fast path can be regarded as an
electronic toll collection lane on turnpike. The name \name is
inspired by this analogy.}, i.e., immediately flushing it to cache bypassing its verification.
Along with the above compiler optimizations, this hardware support can relieve the
pressure on the small store buffer of in-order cores and thus reduce its
structural hazards effectively.
%
%
\ignore{. In summary, \name can (1) lower the number of stores to be
verified by removing unnecessary checkpoints, (2) bypass the store verification, and (3)
reduce the execution delay of the remaining checkpoints, thereby minimizing the performance overhead.
}
Experiments with 36 benchmarks from SPEC2006/2017/SPLASH3 suites
highlight \name's low performance overhead, i.e., 0\% and \revision{14\%} on average---while
Turnstile's overhead is 29\% and 84\%---for 10 and 50 cycles of WCDL, respectively.
Our contributions are below:
\begin{itemize}
    \item \name is the first to make acoustic-sensor-based soft error resilience
        work for in-order cores at a low HW/run-time cost.
    \item We show how compiler optimizations are used not only to remove unnecessary
        checkpoints but also to reduce the execution cycle of the unremoved checkpoints.
    \item We propose 2 novel compiler optimizations to lower the number of registers being checkpointed and reduce register-spilling stores during register allocation.
    \item We propose 2 new hardware schemes to bypass store verification
        without compromising resilience guarantee.
\end{itemize}

\section{Background}\label{sec:background}
This section describes how Turnstile, the state-of-the-art work,
achieves lightweight sensor-based soft error verification with 
region-level error detection and recovery.

\subsection{Region-Level Soft Error Verification}
\label{sec:region-level}
\begin{figure}[h!]
    \centering
    \centerline{\includegraphics[width=\columnwidth]{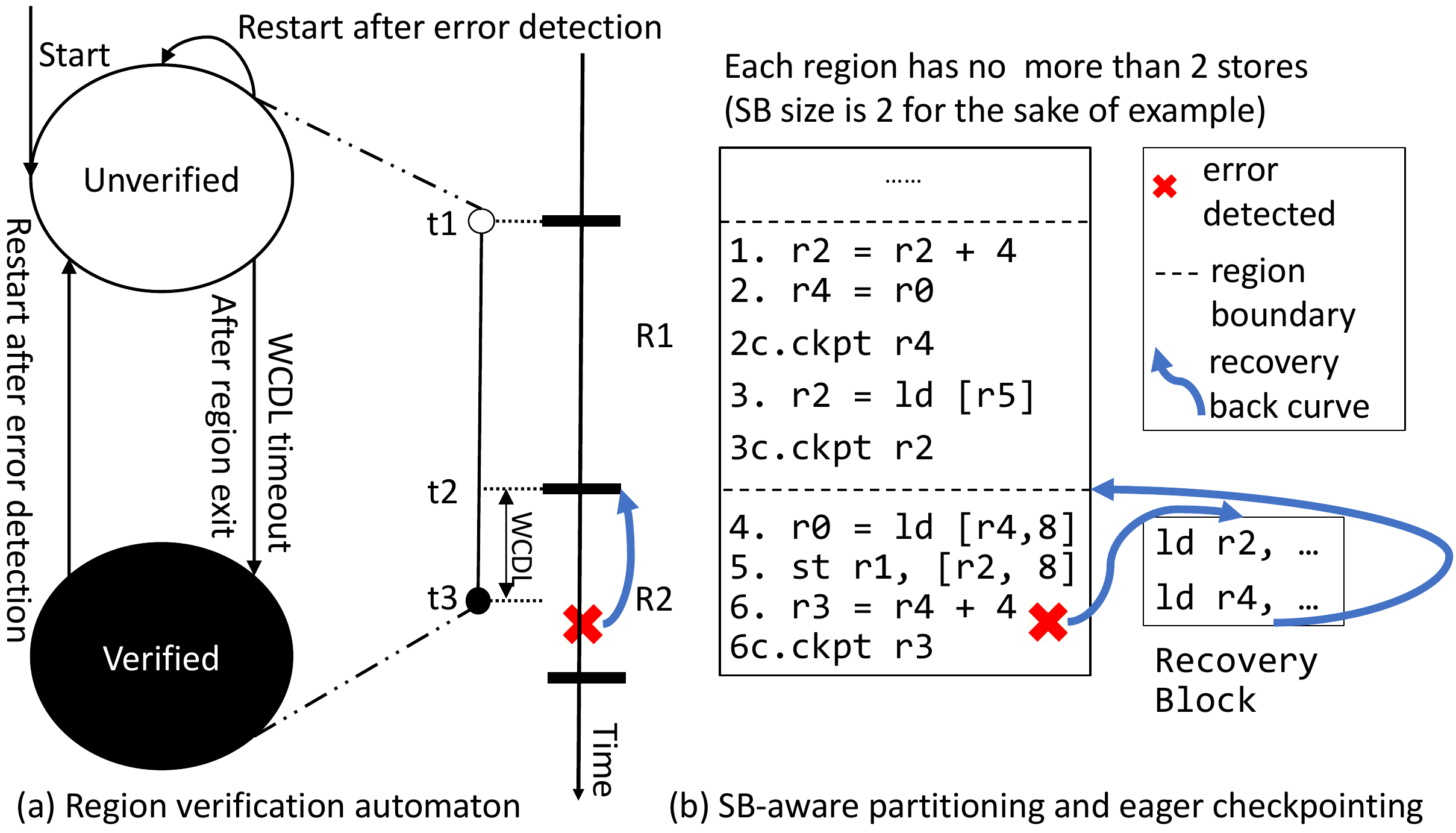}}
    \caption{(a) Turnstile's region verification automaton; (b) store buffer aware region partitioning; eager checkpointing}
    \label{fig:ckpt_instrument}
\end{figure}

To realize the sensor-based soft error verification at a low cost,
Turnstile's compiler partitions the entire program into a series of
verifiable/recoverable regions with the store buffer (SB) in mind so
that each region cannot have more stores than the SB size~\cite{liu2016low}. 
As shown in Figure~\ref{fig:ckpt_instrument} (a), 
each started region is treated as unverified at the beginning, e.g., $R1$ gets $Unverified$ state at time $t1$.
Thus, Turnstile prevents all
the stores of the region from being merged to cache until the region
is verified to be error-free. That is, no sensor detects an error during
the worst-case detection latency (WCDL)---e.g., from $t2$ to $t3$---after the region is finished.
That way Turnstile can contain all the errors occurred during the execution
of a region within the core, keeping cache/memory intact.
Furthermore, this allows Turnstile to correct an error by simply reading
verified data from cache/memory protected by ECC in modern processors
including even low-power cores such as ARM Cortex series.

For the in-core error containment,Turnstile leverages its SB as a gated store
buffer (GSB)~\cite{choi2019cospec,liu2016lightweight}; hereafter, SB refers to GSB. That is, it holds by default all
store write-backs for quarantine even after ROB retires the stores.  To get
them out of the SB quarantine, if verified (i.e., no error detected during WCDL time after the end of their region),
	Turnstile devises a region boundary buffer (RBB) shown in
Figure~\ref{fig:hardware}.  Whenever a region boundary is encountered, i.e.,
one region finishes and the next starts as at $t2$ in
Figure~\ref{fig:ckpt_instrument} (a), Turnstile allocates the RBB entry to
delineate the previously quarantined stores that will be released on their region
verification.  Especially when a region is verified, e.g., $R1$ at $t3$, the
RBB marks the boundary, at which the verified region has ended, as a {\it recovery
PC} in case of a future error.

\begin{figure}[h!]
    \centerline{\includegraphics[width=\columnwidth]{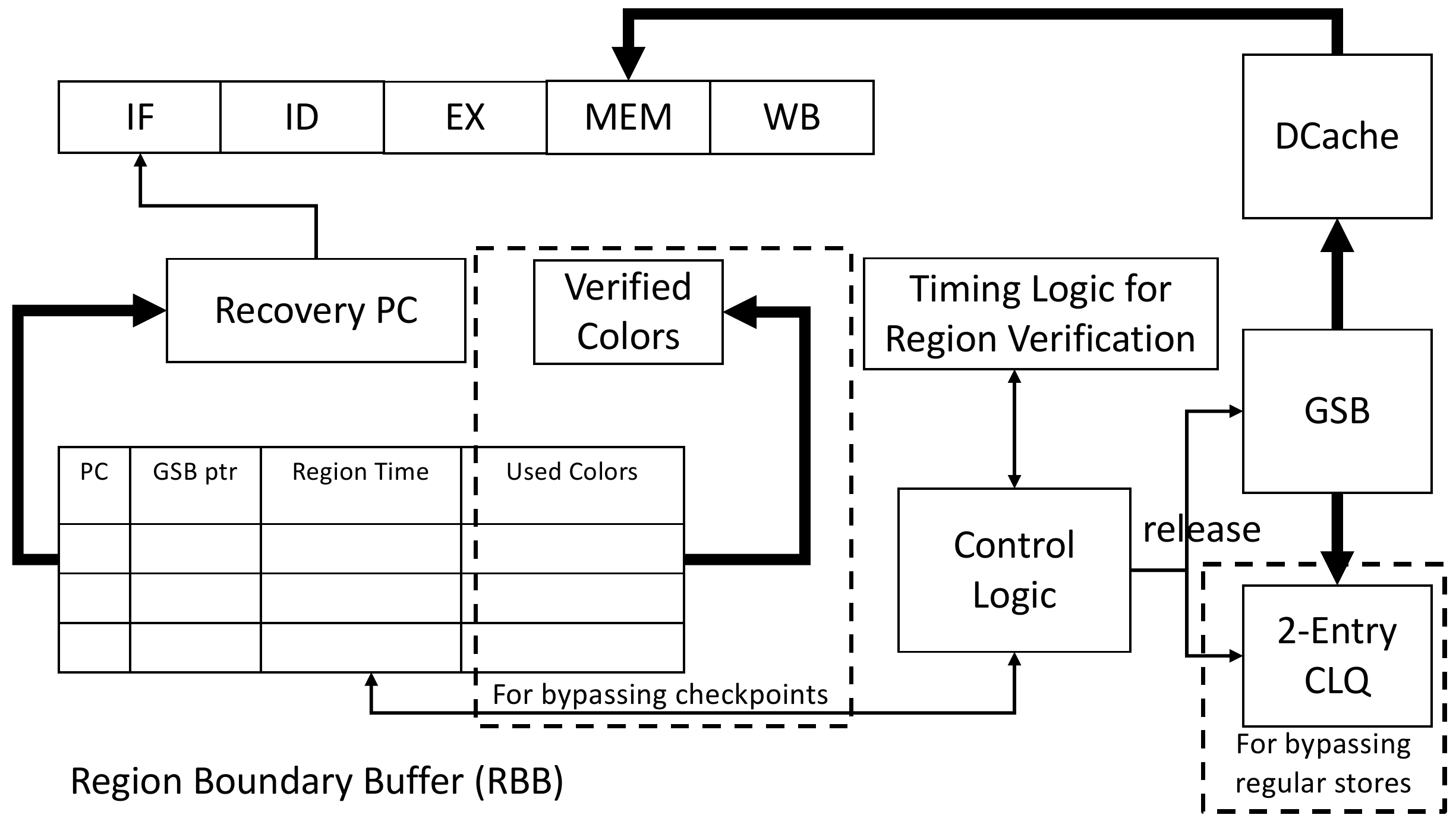}}
    \caption{The high-level view of \textbf{\name}; bold lines correspond to data paths while thin lines to control paths}
    \label{fig:hardware}
\end{figure}

\subsection{Eager Checkpointing and Error Recovery}
The Turnstile compiler performs so-called eager
checkpointing~\cite{liu2016low} that immediately saves updated
live-out registers in memory. That is, it inserts a checkpoint store right after the
register-update instruction provided the register is used as the input of
later regions, e.g., at line 2c, 3c, and 6c in Figure~\ref{fig:ckpt_instrument} (b).  The implication
is three-fold. First, even if a region has multiple updates of a register, only
the last one is checkpointed as the live-out register, e.g., Turnstile checkpoints only the definition
of $r2$ at line 3 though it is pre-defined at line 1 in the figure. Second, Turnstile can
turn register verification into memory verification; registers are verified
through their checkpoint---which is essentially a store instruction---in the
same way as stores are verified, without delaying any register write-back
for performance reasons. Third, the checkpointed register values should be loaded to recover from a soft error.

Upon the detection of a soft error, Turnstile first
discards all SB entries---because they could have been corrupted by the
error---and identifies the most recently verified region boundary by referring
to the {\it recovery PC} and the region starting thereafter.  As shown in Figure~\ref{fig:ckpt_instrument} (b), Turnstile
then executes the recovery block of the region to restore its input (live-in) registers,
e.g., $r2, r4$ in the figure, from the ECC-protected memory (cache) where their
checkpoints have been stored safely with the in-core error containment. Finally,
Turnstile restarts the region recovering from the error. This soft error verification of Turnstile works well for out-of-order cores. However, it incurs a significant run-time overhead for in-order cores as shown in the next section.

\ignore{
\subsection{Acoustic-sensor-based soft error detection scheme}\label{sec:acoustic_detect}
When a high-energy particle strikes the chip, \HJ{an} electron is raised from \HJ{a} ground state to an
excited state. Due to the instability of \HJ{the} excited state, the electron will return \HJdel{back} to a
lower energy level, which induced emits a quantum of energy (such as photon and phonon)
\cite{young1996university}.
By sensing such generated photons and acoustic \HJ{waves}, we can know when and where particle
striking appears. With the help of such \HJ{a} phenomenon, recently, Upasani~\emph{et~al} proposed
a promising technology~\cite{upasani2012setting} to deploy \HJ{the} desired number of sensors on top
of \HJ{the} chip to detect striking in a \HJ{low-cost} manner. For each sensor, it can detect the striking
within 500ns (1000 cycles at 2GHz). Hence, it only requires 300 sensors to achieve 30 cycles
of \HJ{worst-case} detection latency (WCDL) on \HJ{a} 2GHz processor at the cost of less than 1\% chip
area overhead.

\subsection{Fine-grained unified data verification based on sensor-based detection}\label{sec:Turnstile}
Liu~\emph{et~al}~\cite{liu2016low} devised Turnstile to achieve 100\% \HJ{error-free}
in a \HJ{low-cost} manner. First, Turnstile divides the input program into a series of regions
separated with region boundary instruction by taking the store buffer size into consideration
so that Turnstile make sure that the number of store in each region is not greater than a
threshold, for example, half of store buffer size.

After partitioning the program, Turnstile eagerly inserts checkpoint store instruction right
after the last updating point to save the value of a minimal set of live-out register
\footnote{live-out register is that a register is used along some program paths to the program
exit from a specified program point.} to a region to the memory. The minimal set of checkpointed
registers \HJ{are} defined as follows.
$$
    MinCkptSet_{region} = DefReg_{region} \cap LiveOutReg_{region}
$$
Where $DefReg_{region}$ is the set of registers defined in the region and $LiveOutReg_{region}$
is the set of live-out registers to the region. 

{\begin{figure}
    \includegraphics[scale=0.3]{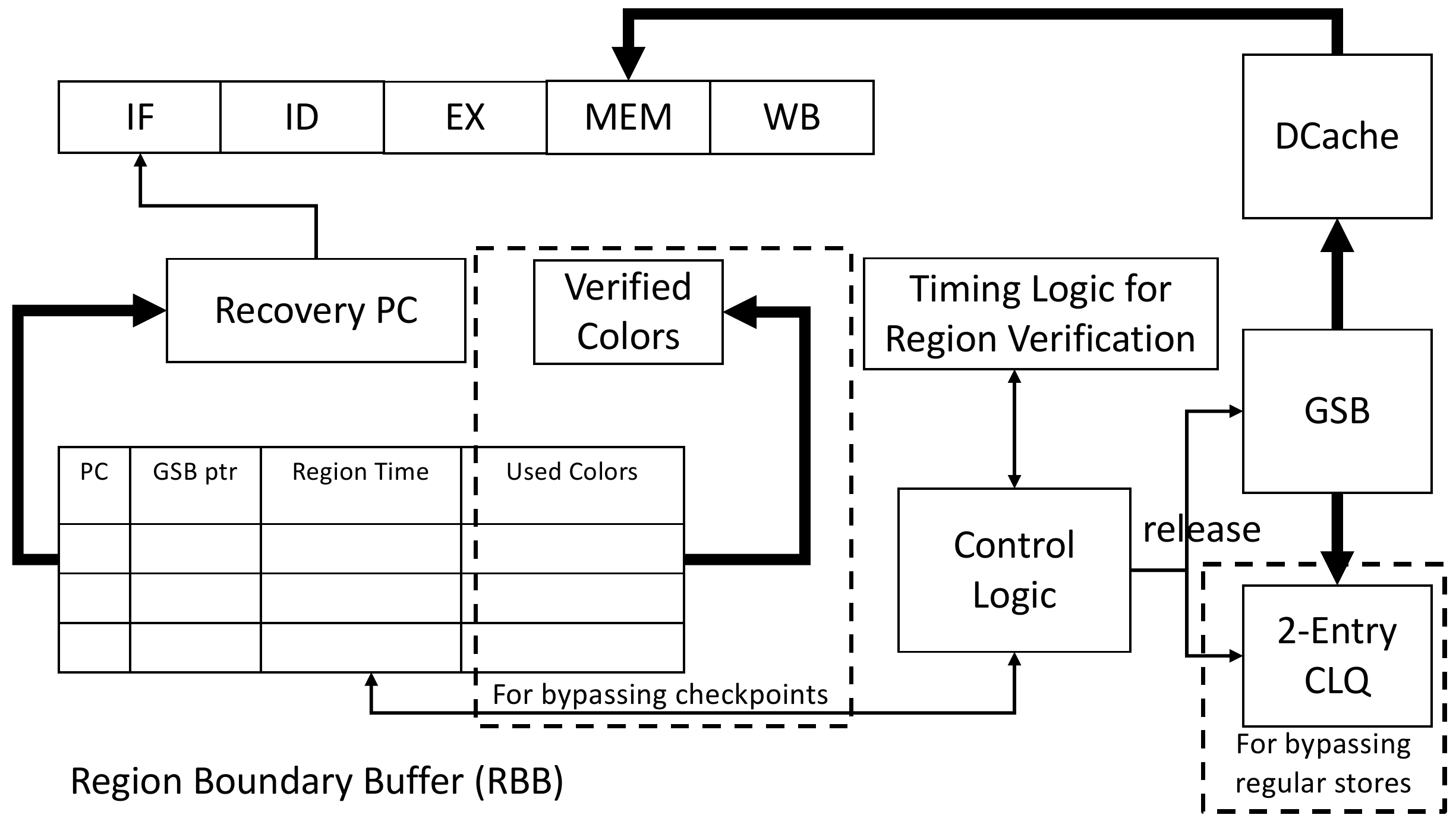}
    \caption{The high level design of \textbf{\name} hardware structure}
    \label{fig:hardware}
\end{figure}}

Figure~\ref{fig:hardware} shows \HJ{the} hardware structure of Turnstile\footnote{The extra hardware
components added for our purpose is highlighted with dotted lines.}. The region boundary
buffer (RBB) is a crucial hardware \HJ{component} for maintaining necessary information about
the regions which are being verified. GSQ means gated store buffer which is used for buffering
all \HJ{stores} being verified and controlled by \HJ{the} control logic to release some stores once the
region is verified. For each region, there is \HJ{an} RBB entry allocated in the RBB when \HJ{the} pipeline
encounters a region boundary instruction. Each RBB entry has three fields, PC offset of region
boundary instruction leading that region, GSQ ptr which points to the tail of store buffer
when that region boundary is encountered, Region Time representing the execution time of last
region ending with that region boundary. There are two auxiliary counters, To Waited and Has
Waited, to facilitate releasing of \HJ{the} oldest region corresponding to the head RBB entry. Once the
counter To Waited becomes zero, the region corresponding to the head RBB entry is verified
successfully. Turnstile saves the PC value of head RBB entry to the Recovery PC register.
Subsequently, Turnstile tells control logic to release head RBB entry and all store instructions
being kept in store buffer before the position pointed to by GSQ pointer. If the sensor-based
detection generates a signal that soft error presents, Turnstile simply drains the gated store
buffer, performs the recovery logic to re-construct desired register values from checkpoints,
and resets the PC register with the value of Recovery PC register, then redirect the program
control flow to the latest verified region boundary to restart the program.
}

        \vspace{-5pt}
\section{Motivation}\label{sec:problem}
This section discusses 3 main reasons why Turnstile, the state-of-the-art work, incurs a high 
run-time overhead for in-order cores: (1) Turnstile's checkpoints put significant pressure on the small store buffer (SB) leading to the
structural hazards. (2) Once an SB entry is allocated, it stays long therein
 till the region is verified, which keeps holding the pressure 
and makes it take a while to resolve the structural hazards.
(3) Due to eager checkpointing, the dependence between a register-update
instruction and its immediate successor (i.e., a checkpoint store) often
causes data hazards, slowing down the region execution. It is worth noting that the above problems are not a big deal for out-of-order cores
thanks to the large SB ($\geqq$40) and the ability to schedule independent instructions to address
the hazards. In contrast, the problems are devastating for in-order cores in
that the SB has only a few entries (e.g., 4 in ARM Cortex-A53),
and the hazards freeze all following instructions because of the in-order pipeline
execution.

\subsection{Checkpoint Puts Pressure on Store Buffer}\label{sec:excessive_checkpoint}
\begin{figure}[h!]
    \begin{minipage}{0.52\columnwidth}
    \centering
    \includegraphics[width=1\columnwidth]{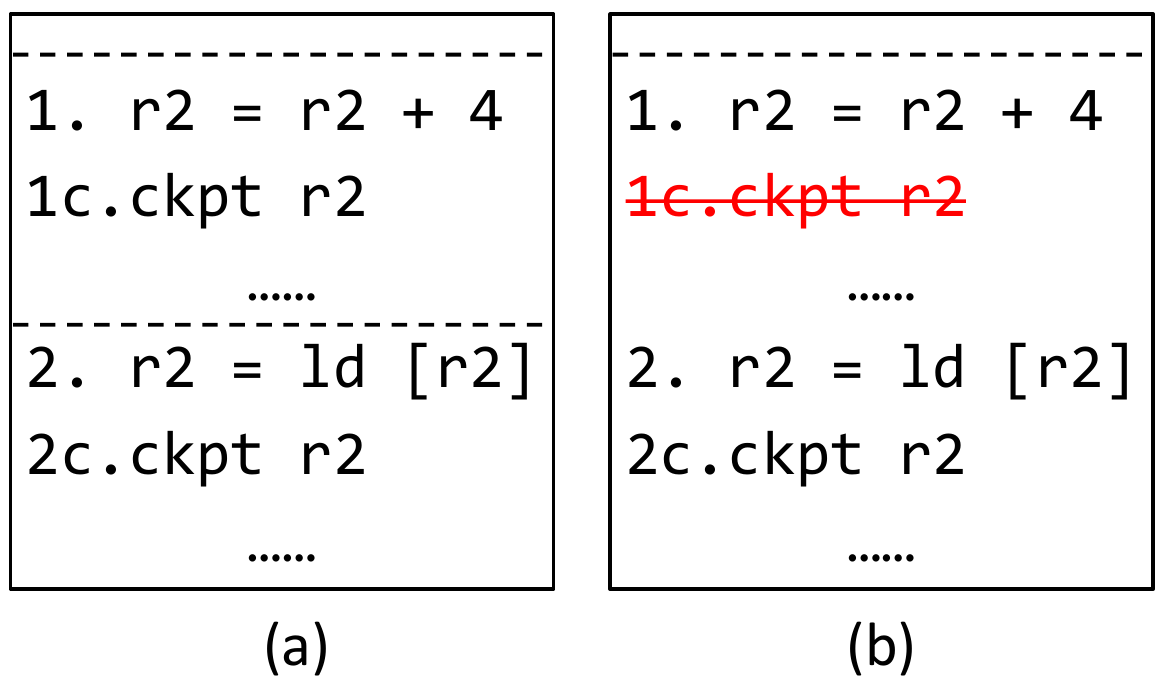}
    \caption{Impact of region size}
		\label{fig:sb_size_impact_on_partitioning}
    \end{minipage}
    \hfill
    \begin{minipage}{0.47\columnwidth}
    \centering
    \centerline{\includegraphics[width=1\columnwidth]{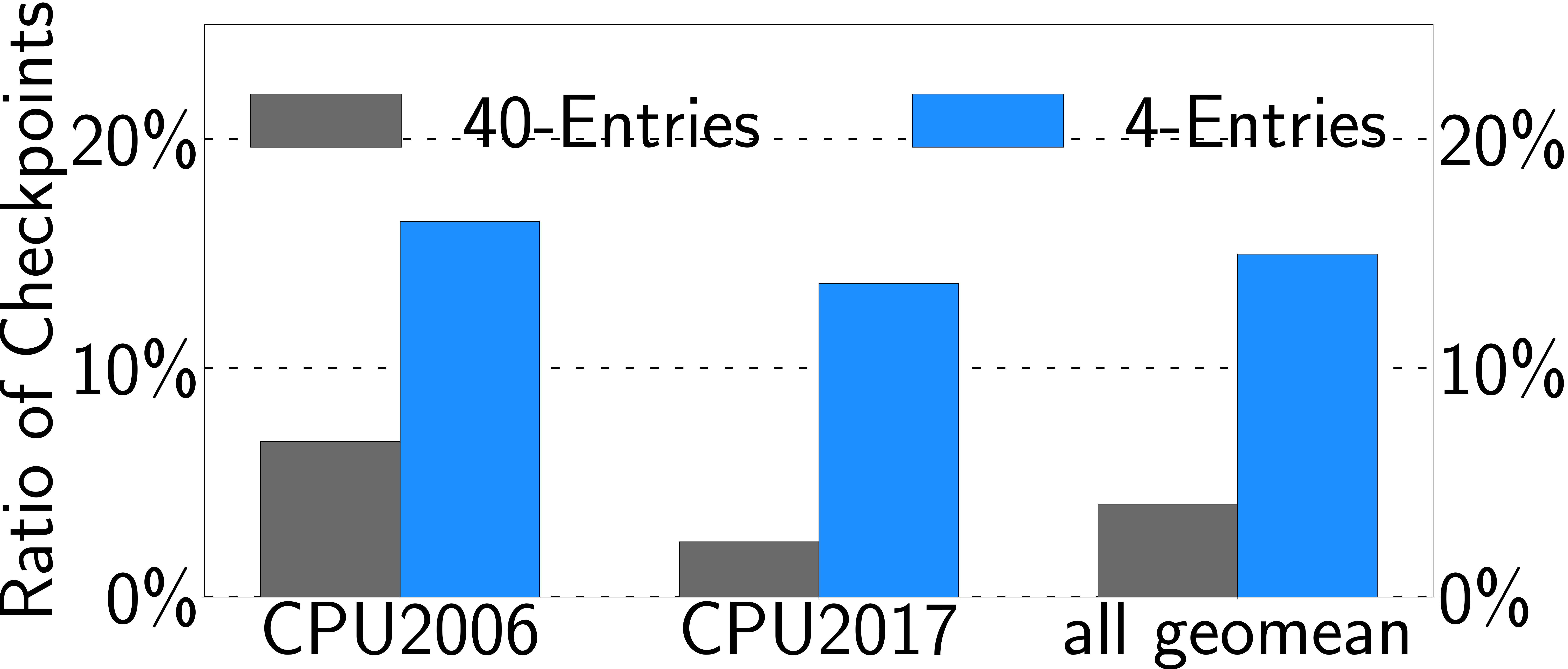}}
    \caption{Impact of store buffer size on the number of inserted checkpoints}
    \label{fig:comparison_sb}
    \end{minipage}
\end{figure}
To equip the verifiable regions---partitioned with the SB size in mind---with the recoverability, Turnstile checkpoints live-out
register values of each region by logging them to memory (Section~\ref{sec:background}).
Since the regions are generally short due to the tiny SB (only 4 entries in modern
in-order cores), the short regions tend to have more live-out registers updated overall than the long
regions for a large SB. For example, while a register $r2$ is live-out in both regions
 and thus checkpointed twice at line 1c and 2c in Figure \ref{fig:sb_size_impact_on_partitioning}
(a), it is checkpointed only once in the same code that does not have a region boundary in-between as shown in
Figure \ref{fig:sb_size_impact_on_partitioning} (b); that is because $r2$ defined at line 1 is no
longer live-out since it is overwritten by the following definition at line 2. 

Figure~\ref{fig:comparison_sb} confirms that the number of inserted checkpoints 
(i.e., store instructions logging the live-out registers)
significantly increases when the store buffer is shrunk from 40 to 4 entries.
When the store buffer (SB)
size is 40 as in out-of-order cores, Turnstile's eager checkpointing accounts
for 4.1\% of the total dynamic instruction count on average for SPEC 2006/2017 benchmark applications.
On the other hand, when the SB size is 4 as in in-order cores, the ratio
significantly increases to 14.98\%. It turns out that such many checkpoints often fill
up the SB, making the next store stall the pipeline due to the lack of room in
the SB, i.e., the structural hazard. 
In particular, we found it possible to
remove many of the checkpoints without compromising the soft error resilience;
Section~\ref{sec:chkpt-removal} discusses it in detail.


        \vspace{-5pt}
\subsection{Verification Keeps the SB Pressure Long}\label{sec:verification_delay}
\begin{wrapfigure}{R}{0.45\columnwidth}
        \centering
        \centerline{\includegraphics[width=0.45\columnwidth]{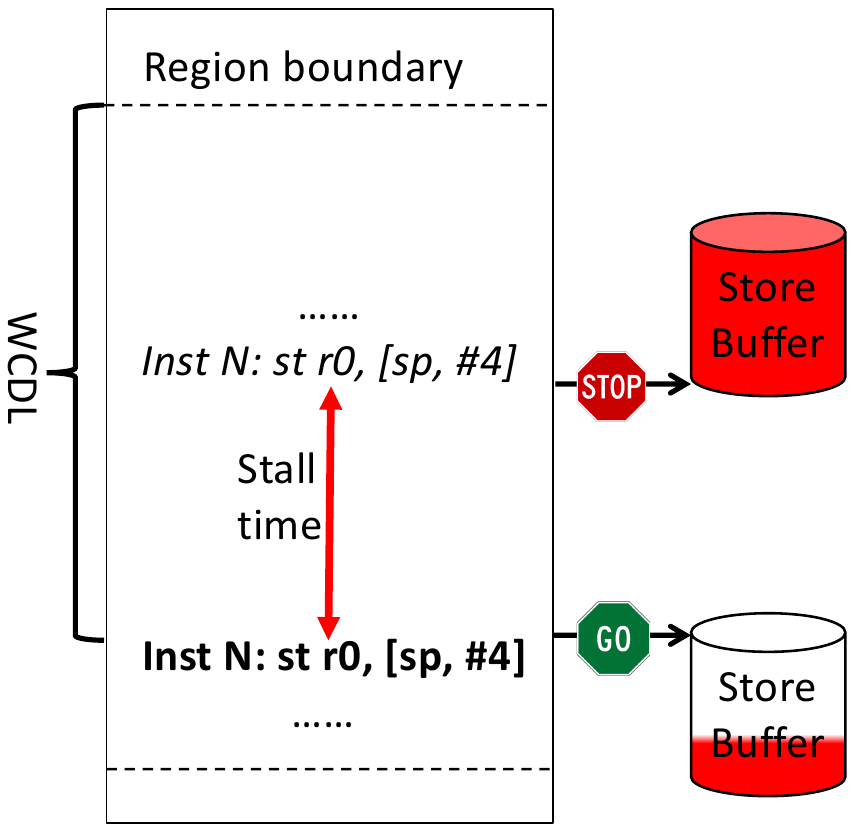}}
        \caption{Stall due to the lack of room in the SB}
        \label{fig:verification_delay}
\end{wrapfigure}

To verify each region (ensuring the absence of soft errors during the
region execution), Turnstile holds all the data being stored in the SB
till the region is verified to be error-free.  Hence, no allocated SB
entries of stores can be released to L1 cache during the execution of their
region; rather, they can only be released WCDL (10-50) cycles later
after the region ends.  The implication is that stores cannot but reside
in the SB for such a long period of verification time, keeping the high
pressure on the SB.
In essence, this may cause a structural hazard if the SB has already been
full when the pipeline encounters a new store, e.g., \texttt{inst N:
st} in Figure~\ref{fig:verification_delay}. Unfortunately, the hazard 
cannot be resolved until the prior region is verified with its stores released to cache.
In other words,  the pipeline stall continues all the way to 
the region verification point---where the WCDL time elapses in
the figure. Here, due to the in-order nature of the pipeline, it cannot
schedule any of the following instructions thus being unable to hide such a long stall latency.
As such, it postpones not only the stalled store instruction, e.g., \texttt{inst N: st} in the figure, but also all the subsequent
instructions. 
As will be shown in Section~\ref{sec:eager_release}, \name can safely 
release some stores from the SB without holding them for verification, thereby relieving the store buffer pressure.

\subsection{Eager Checkpointing Slows Down the Store}\label{sec:execution_delay}

Turnstile's eager checkpointing introduces a read-after-write dependence
between the instruction, that updates a live-out register, and its checkpoint
store. This is particularly harmful for the in-order pipeline because of the
inability to dynamically schedule other independent instructions---unlike
the out-of-order pipeline that can overlap their execution with the live-register-update instruction. 
In an in-order core, checkpoint stores can often be stalled since the register being checkpointed is not available for their execution.

\begin{figure}[!htb]
    \centering
    \centerline{\includegraphics[width=1.0\columnwidth]{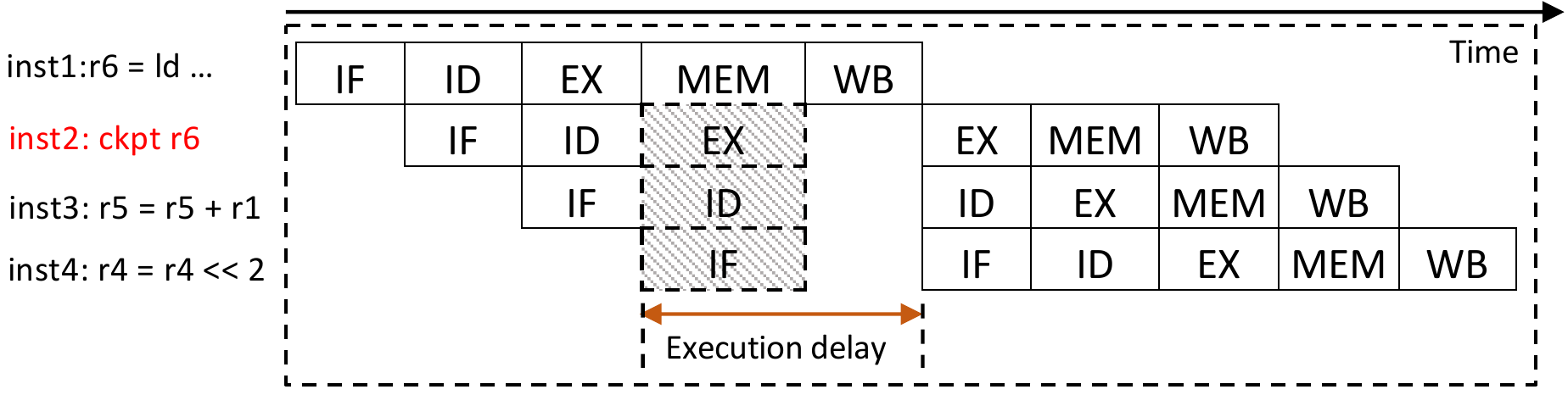}}
    \caption{Checkpoint's execution delay on in-order pipeline}
    \label{fig:execution_delay}
\end{figure}

In such a case, the in-order pipeline must be delayed for a certain time to resolve the
data hazard, e.g., till the value of register $r6$ becomes available in Figure~\ref{fig:execution_delay}
where checkpoint store is marked in red. Since it is the load instruction that updates the
$r6$ in this example, the execution delay of the checkpoint store could be
significant on cache misses. The takeaway is that such a checkpoint execution delay translates
to the significant extension of the program execution time;
Section~\ref{sec:inst_sched} shows how \name reduces the delay to make the checkpoint store instruction execute faster.

\section{\name for Solving the Problems}\label{sec:overview}

\begin{figure}[h!]
    \includegraphics[scale=1, width=1\columnwidth]{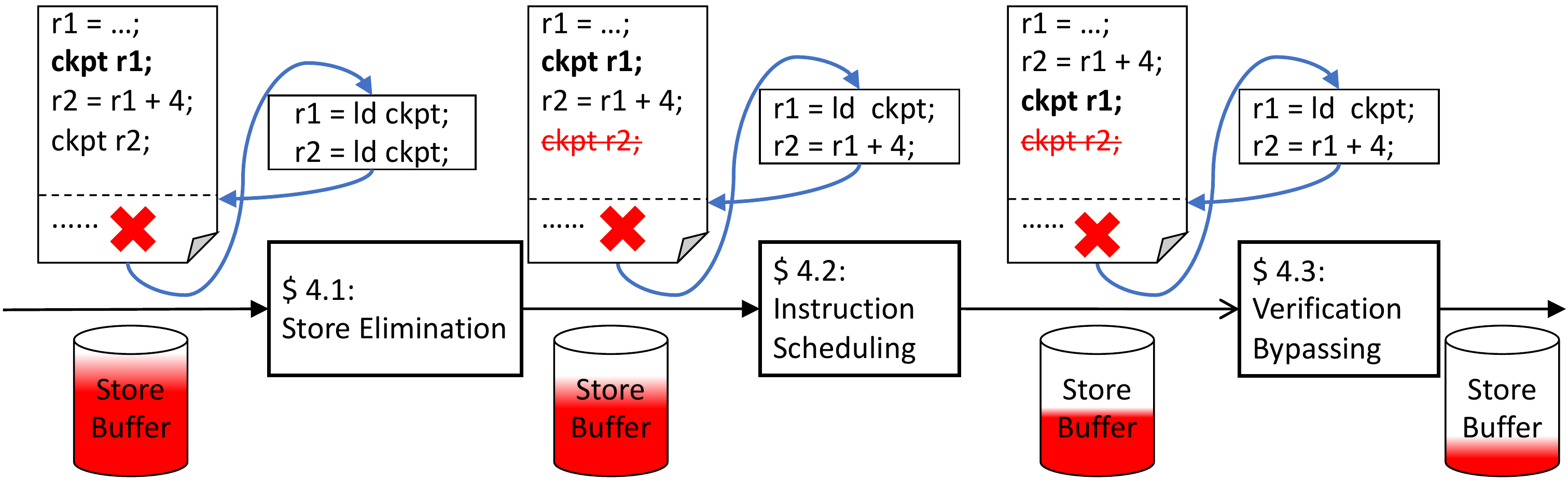}
    \caption{The 3 phases of \name SW/HW optimizations}
    \label{fig:overview_opts}
\end{figure}
To address the 3 problems in Section~\ref{sec:problem}, our proposal, \name, leverages 
compiler and architectural optimizations; Figure \ref{fig:overview_opts} shows the
workflow of the 3 optimization phases.

In the first phase (Section~\ref{sec:chkpt-removal}), \name's 2 new compiler optimizations reduce the traffic to store buffer by (1) generating less spilling stores
during register allocation and (2) eliminating a loop induction variable being checkpointed if it can be merged with others.
Likewise, \name removes unnecessary checkpoints with two existing compiler optimizations,
checkpoint pruning~\cite{liu2016compiler} and loop invariant code motion (LICM)~\cite{muchnick1997advanced} to further lower the store buffer traffic. 
Second, for the remaining checkpoints that cannot be removed by the first phase,
the \name compiler performs checkpoint-aware instruction scheduling to hide
the delay caused by data hazards (Section~\ref{sec:inst_sched}). 
Finally, to directly reduce the pressure on the store buffer (SB), \name leverages 2
novel hardware techniques---in Section~\ref{sec:eager_release}---that can (1) skip the verification of the remaining
checkpoint stores and the regular stores and (2) merge them to cache right after
they are committed.  
Note that the above 3 optimization phases have a synergistic impact on
reducing the SB pressure. 
The rest of this section details the 3 optimizations.

\subsection{Reducing the Traffic to the Store Buffer}
This section shows how to reduce the SB traffic with 2 new compiler optimizations and 2 other existing ones.
While the first addresses regular stores, the next 3 optimizations do checkpoint stores.

\label{sec:chkpt-removal}

\subsubsection{Store-Aware Register Allocation}\label{sec:ra_trick}
In addition to application stores, the other source of regular stores is register allocation.
Since it is done in a best effort manner, some variables end up being spilled to stack memory when architectural registers run out during the
register allocation. To avoid spilling performance-critical variables,
				 traditional register allocators takes a heuristic approach to
				 determine what to spill. More precisely, they maintain the spill
				 cost (weight) of variables which summarizes the execution frequency of
				 their use points (reads and writes).
Unfortunately, since the spill code models of traditional register allocators
do not differentiate writes from reads, they may generate superfluous spilling
stores. While this is not a concern for most processors where stores are off
the critical path, in-order cores equipped with sensor based soft error
verification can suffer a significant performance degradation.  To address this
problem, the \name compiler increases the cost for the write operation of each
variable in the spill candidate decision logic. Note that care must be taken to 
maintain the original register allocation quality in terms of the number of spilled variables, which would otherwise degrade the performance of the resulting code. As a result, \name can keep those
variables, that are frequently written, in architectural registers, and thus
all the writes to the variables just become register writes other than memory
stores. 

\subsubsection{Loop Induction Variable Merging (LIVM)}\label{sec:indvar_deopt}

\begin{figure}[h!]
    \center
    \includegraphics[width=1\columnwidth]{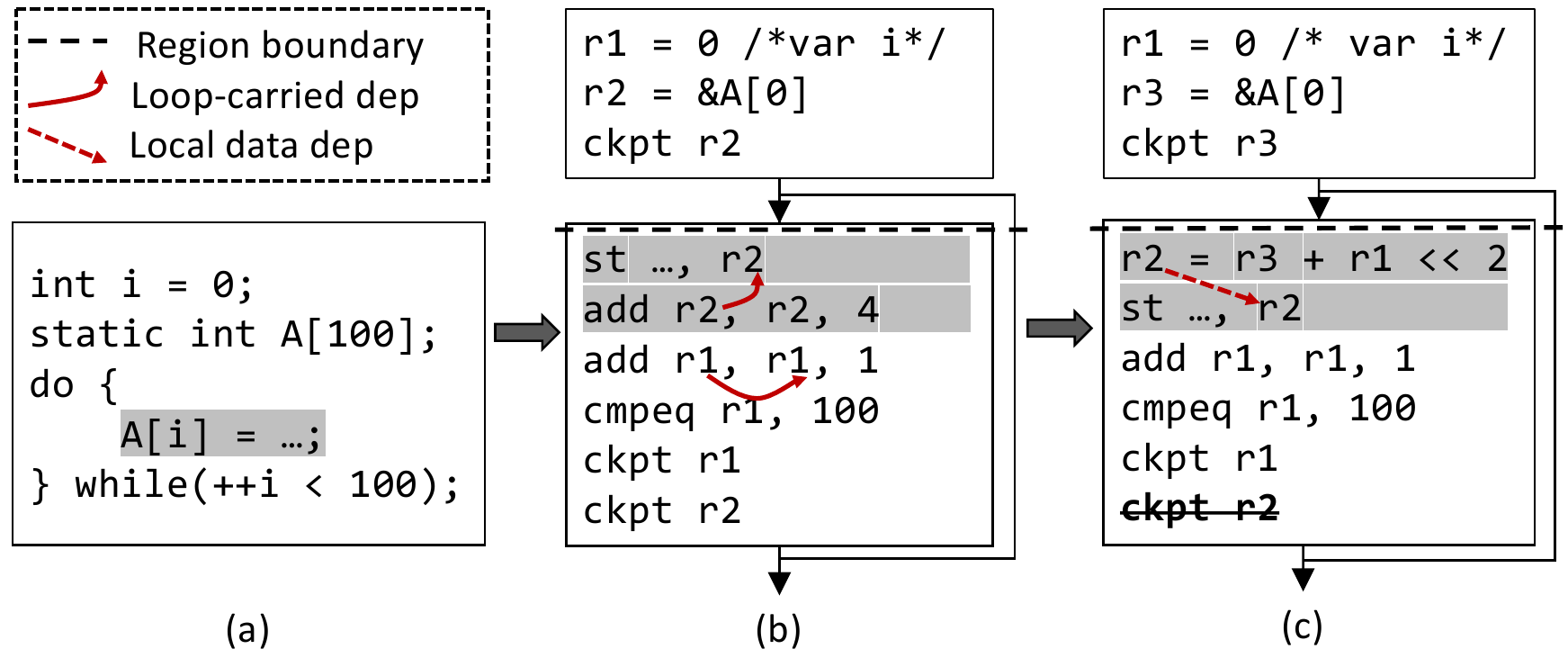}
    \caption{(a) original C code, (b) strength reduction code, and (c) LIVM enabled code eliminating $r2$'s checkpoint}
    \label{fig:indvar_merge}
\end{figure}
We found that traditional compilers often generate additional loop
induction variables---that must be checkpointed each loop iteration---and the
resulting checkpoint stores increase the store buffer traffic significantly.
There are two kinds of induction variables~\cite{muchnick1997advanced}: basic induction variable, e.g.,
$i$ in Figure \ref{fig:indvar_merge} (a) and induced induction variable whose value is a (linear) function of a basic induction
variable, e.g., the address expression of $A[i]$ in Figure \ref{fig:indvar_merge} (a).

Here, the compiler's loop strength reduction~\cite{muchnick1997advanced} turns the expression ($\&A[0]+i*4$) into a separate basic induction variable that is initialized as $\&A[0]$ and increased by 4 as shown in Figure \ref{fig:indvar_merge} (b).
The problem is that the strength reduction results in loop-carried data dependence,
i.e., $r2$ is used in the next iteration as the address operand of a store, rendering $r2$ live-out~\footnote{A region boundary is placed in a loop header as in Turnstile~\cite{liu2016low}.} and checkpointed in the loop thus degrading the performance; Figure~\ref{fig:indvar_merge} (b) highlights the strength reduction enabled code in the shaded box and shows the resulting checkpoint, i.e., $ckpt~r2$, in the bottom.

To deal with the problem, \name proposes a new optimization called loop
induction variable merging (LIVM). It investigates basic induction variables in
a loop to see if one can be merged to some other basic induction variable in
form of an expression derived from the basic induction variable. In other words,
LIVM makes the merged variable become an induced induction variable so that it can eliminate the loop-carried data dependence. 
As shown in Figure~\ref{fig:indvar_merge} (c), $r2$ has only local data dependence with the help of LIVM; since $r2$ is no longer live-out, \name eliminates the 
 $ckpt~r2$ in the bottom of the figure. Since it used to be executed every loop iteration, the impact of its elimination on the store buffer traffic reduction should be very significant if enabled.

\subsubsection{Optimal Checkpoint Pruning}\label{sec:pruning}
To further reduces checkpoint stores, \name leverages optimal checkpoint pruning~\cite{pennyPLDI20} in the recent
advance of GPU register file protection called Penny. We found the
pruning technique effective for reducing the store buffer pressure, though it
is originally devised for GPUs---that have never had a store buffer (SB)---and
idempotent regions~\cite{pennyPLDI20} that are intrinsically different from
\name's SB-size aware partitioned regions.

\begin{figure}[h!]
    \center
    \includegraphics[width=0.7\columnwidth]{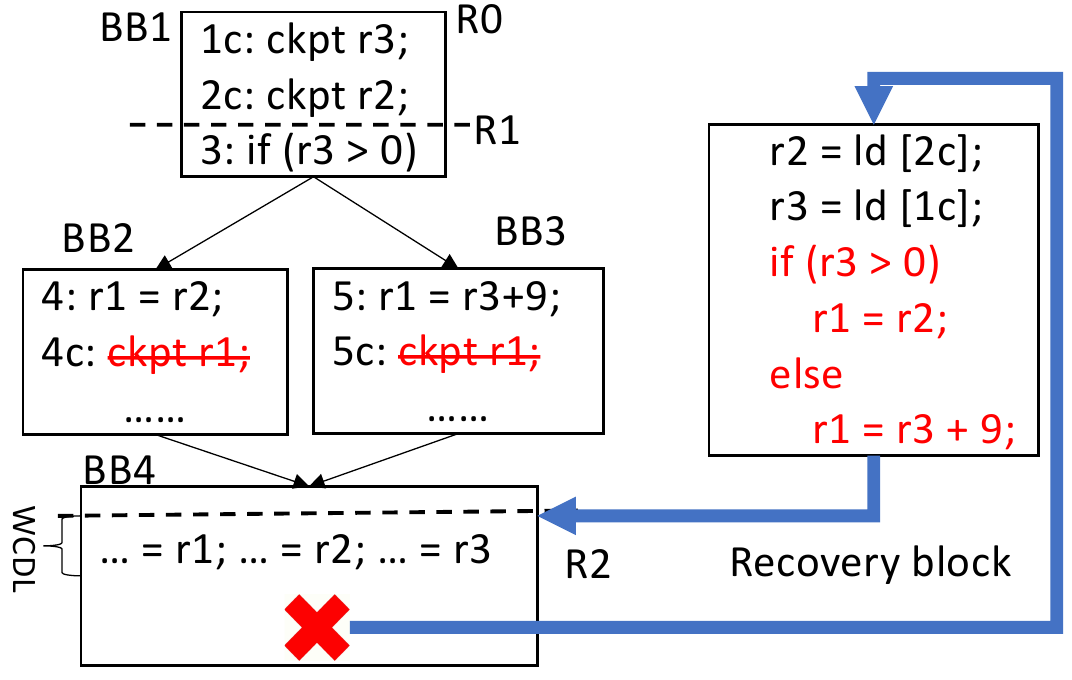}
    \caption{Checkpoint pruning for eliminating $4c$ and $5c$}
    \label{fig:ckpt_elimination}
\end{figure}

It turns out that Penny's checkpoint pruning removes a large number of
checkpoints without compromising the recoverability guarantee.  The key idea is
that it is safe to remove those checkpoints, provided the value to be
checkpointed can be reconstructed from a constant or the value of other
checkpoints at the recovery time of an error detected.  \name exploits Penny's optimal pruning
algorithm that can detect unnecessary checkpoints in polynomial time with the
recovery code generated to reconstruct the pruned checkpoint value.

Figure~\ref{fig:ckpt_elimination} shows how the checkpoint pruning works and ensures
safe recovery. Suppose an error is detected in a region R2 in the bottom of
the figure; R2's input registers $r1, r2, r3$ have been checkpointed by prior
regions, e.g., the first region R0 checkpoints $r2$ and $r3$. Similarly, without
the checkpoint pruning, $r1$ would be checkpointed by the middle region R1
where either $r2$ or $r3$ is used to update $r1$ depending on the path taken in
the branch. Here, either way, $r1$'s checkpoint ($4c$ and $5c$) can be removed
since it can be reconstructed by using the checkpointed value of $r2$ or $r3$
in the recovery block. To recover from the error here, the recovery block of
the region R2---starting from the recovery PC
(Section~\ref{sec:background})---executes the backward slice of the pruned
checkpoint, which includes the branch to reconstruct $r1$ differently according
to the checkpointed predicate $r3$, and jumps back to the recovery PC, i.e.,
	 the beginning of the region R2.

\subsubsection{Moving a Checkpoint out of a Loop with LICM}
\begin{wrapfigure}{R}{0.55\columnwidth}
    \centering
    \includegraphics[width=0.55\columnwidth]{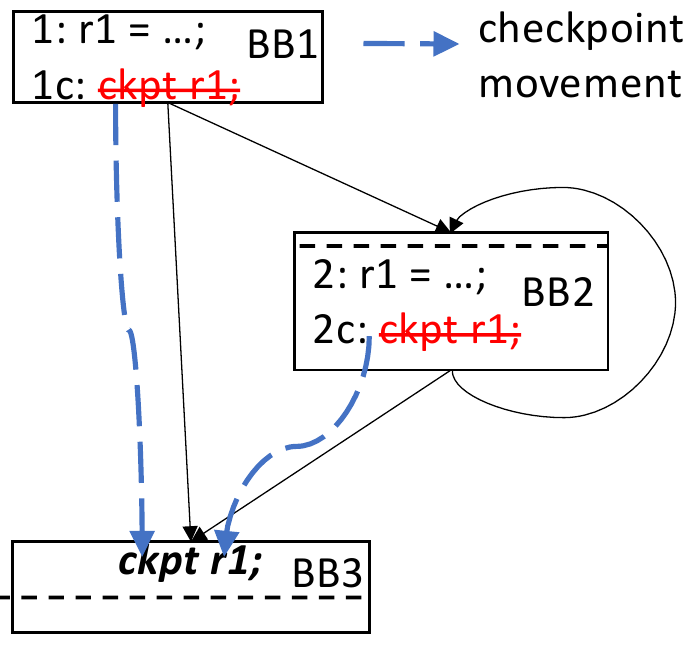}
    \caption{LICM at $2c$}
    \label{fig:licm}
\end{wrapfigure}
Although the pruning scheme investigates if a checkpoint can be safely eliminated,
it never tries to move the location of the checkpoint.  For those checkpoints
that cannot be eliminated, the pruning scheme leaves them at their original
checkpointing location, thus losing a chance to move a checkpoint out of the
loop body. The main reason for this is that under the eager checkpointing
policy, a checkpoint cannot but be placed right after the instruction that
updates the live-out register in each region.

Interestingly, the eager checkpointing can be relaxed without compromising the
recoverability. Recall that a checkpoint is necessary for 2 reasons: (1) saving the registers
that are input to some later regions and (2) verifying the integrity of the
register value, i.e., no register corruption. In the input-saving point of
view alone, \name only has to checkpoint the register before it is used. On the
other hand, to verify the register, it must be saved before the region is
finished due to the region-level verification (Section~\ref{sec:region-level}).  As a result, for each
checkpoint in a given region, the checkpoint can be safely moved from the
original eager checkpointing location down to any points before the region boundary.

Figure~\ref{fig:licm} shows how \name leverages this insight to move $r1$'s
checkpoint at line $2c$ out of a loop as in LICM (loop invariant code motion\footnote{
While LICM is to hoist the invariant code out of a loop~\cite{aho1986compilers,cooper2011engineering},
most of the production compilers (GCC/LLVM) have extended LICM to support
code sinking too. We modified the LLVM passes to
move down checkpoints in a loop as they are guaranteed not to be aliased
with other memory operations.}).  
By moving the checkpoint down to near the region boundary below,
\name takes the checkpoint off from the loop body. 
Moreover, since the checkpoint is now placed in the bottom basic block, another
checkpoint at line $1c$ becomes redundant---as they both checkpoint the
same value of $r1$---and can be safely eliminated as well.  That way \name
can reduce the performance overhead significantly for some applications (Section \ref{sec:optimizations_effect}).

\subsection{Hiding the Execution Delay of Checkpoints}\label{sec:inst_sched}
There are still many remaining checkpoints that cannot be removed by the prior
2 optimizations. Since Turnstile inserts each checkpoint store right after the
register-update instruction, the store's dependence on the register (data
hazard) often makes the in-order pipeline stall till the register gets ready
(see Section~\ref{sec:execution_delay}). To address this problem, \name
leverages another compiler optimization, i.e., instruction
scheduling~\cite{cooper2011engineering,aho1986compilers,muchnick1997advanced}.
It attempts to separate the register-update instruction from the dependent
checkpoint store instruction by hoisting some of the following independent
instructions beyond the store instruction.

\begin{figure}[h!]
    \centering
    \includegraphics[width=1.0\columnwidth]{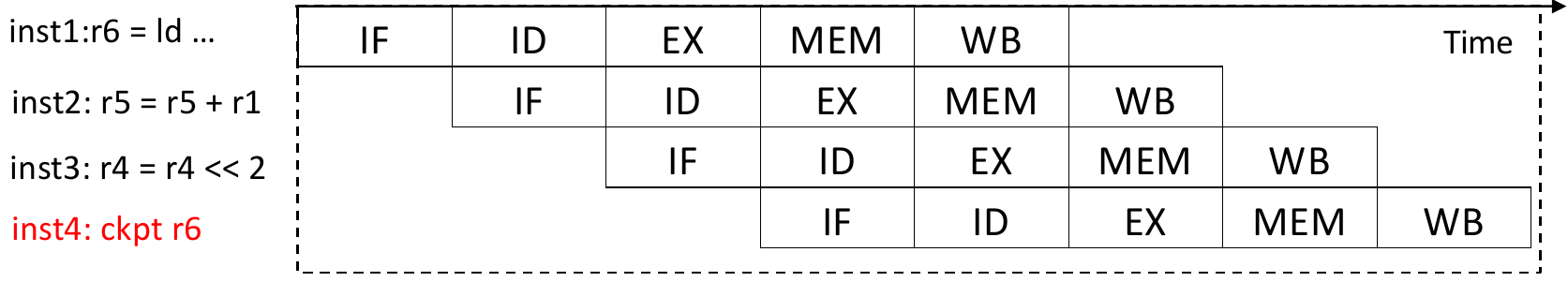}
    \caption{Execution delay of checkpoint gets reduced by rescheduling instruction stream}
    \label{fig:isched}
\end{figure}

Figure \ref{fig:isched} shows how the instruction scheduling can handle the
checkpoint data hazard in Figure \ref{fig:execution_delay}.  With the scheduling,
the checkpoint store for a register $r6$ being loaded is moved down in
Figure \ref{fig:isched}; that way the store can be executed with no stall,
i.e., its operand $r6$ is ready from the load---because the load latency is
overlapped with the execution of 2 other intervening instructions
before the store.  Since the register becomes available when the reordered
store is about to execute, it can avoid the data hazard. 
Note that the instruction scheduling helps the pressure on the
store buffer (SB) to be relieved as well. The reason is that the reordered stores can
eventually reduce the execution time of their region, which is the part of the
region verification time; it consists of the region execution time and the WCDL
as shown in Figure~\ref{fig:ckpt_instrument}(a). Hence, this also reduces
the time during which stores stay in the SB for verification, keeping the
pressure for a shorter amount of time.

\subsection{Relieving the Store Buffer Pressure}\label{sec:eager_release}
\begin{wrapfigure}{R}{0.45\columnwidth}
    \centering
    \includegraphics[scale=0.2,width=0.4\columnwidth]{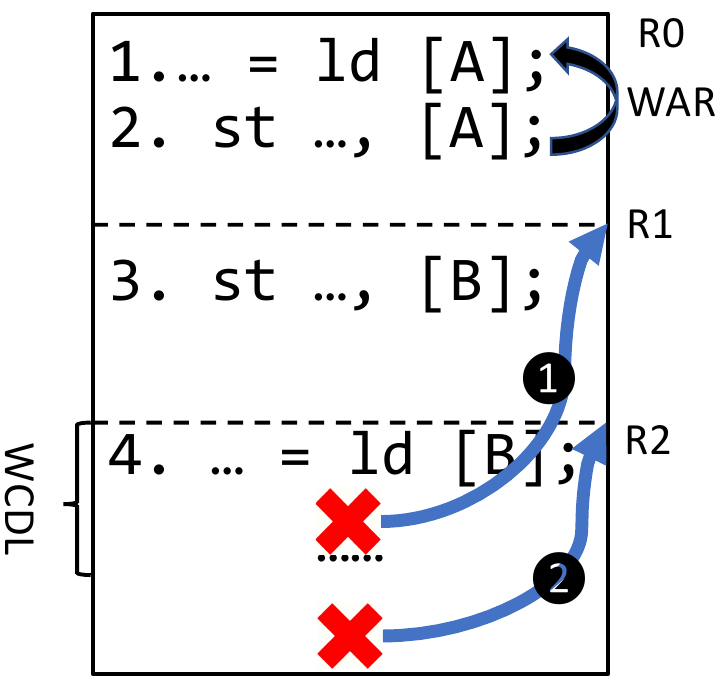}
    \caption{Fast release of a WAR-free regular store}
    \label{fig:anti_dep_checking_motivation}
\end{wrapfigure}

Unlike other optimizations, the next two novel hardware schemes in this section can
directly relieve the store buffer pressure. \name achieves that by releasing
some of the buffered stores to cache without verification yet in a manner that
still guarantees the soft error resilience.  To begin with, we classify store
instructions into two kinds: (1) regular stores stemming from the program itself or
register allocation, i.e., spill stores to stack, and (2) checkpoint stores generated to save updated
live-out registers. The rest of this section discusses the two kinds
and how they are addressed by our 2 hardware schemes, respectively.

\subsubsection{Fast Release of Regular Stores} 
Prior work, Turnstile, has all stores of a region quarantined in a store buffer
(SB) till the region turns out to be error-free for both region
verification and in-core error containment purposes (Section~\ref{sec:background}).
However, we found out that not all the data being stored are going to be read
for the verification of a region. For example, the data stored at line 3 in
Figure~\ref{fig:anti_dep_checking_motivation} is never read in the region R1 when
R1 is restarted upon an error due to the absence of write-after-read (WAR)
dependence; we refer to such a store as a WAR-free store.  Thus, even if the
data is corrupted due to an error and written to cache, R1's re-execution can
correctly recover from the error.  With that in mind, \name releases such a
WAR-free store---without verification---immediately after its commit, thereby
relieving the store buffer pressure.

One might suspect that due to the fast release of unverified data to cache,
the next region might read it making the error recovery fail, e.g., data stored
at line 3 by a region R1 is loaded at line 4 by the following region R2 in
Figure~\ref{fig:anti_dep_checking_motivation}.  Fortunately, it turns out that
this is not a problem at all.  For the unverified yet corrupted data to be read
by the region R2's load, the error must be detected before R1's verification point, i.e.,
within WCDL (e.g., 10) cycles after the prior region R1 is finished.  However,
it is not R2 but R1 that the original region-level soft error verification
(Section~\ref{sec:background}) restarts to recover from the error (\ding{182}
in Figure~\ref{fig:anti_dep_checking_motivation}).  Again, R1 does not read
the data, i.e., no WAR dependence, and therefore restarting R1 can correct the error
with no harm.  On the other hand, if the error is detected after R1's
verification point, then R2 is restarted for recovery (\ding{183} in
Figure~\ref{fig:anti_dep_checking_motivation}). In this case, R2's load is
guaranteed to read the correct data---because it was written by the region R1
which has already been verified.

The takeaway is that WAR-free stores can bypass the verification and thus can
be immediately merged to cache after their commit, whether the error is
detected during the execution of their region or within WCDL cycles after the
region is finished.  To realize this, \name proposes a novel microarchitectural
technique called committed load queue (CLQ)---shown in
Figure~\ref{fig:hardware}---to dynamically check the absence of WAR dependence
for each regular store.

\paragraph{Ideal CLQ Design with Address Matching}
For each committed load, \name allocates an entry in the CLQ to keep the address
of the load.  When the in-order pipeline tries to commit a regular store, \name
compares its address to all the entries of CLQ to check whether the store has
WAR dependence on any prior load in the current region. If there is no address
conflict, i.e., no WAR dependence, \name releases the WAR-free store
immediately instead of holding it in the store buffer.  Otherwise, it is
quarantined in the store buffer as is for the original region-level
verification. 

\begin{figure}[h!]
    \centerline{\includegraphics[width=1.0\columnwidth]{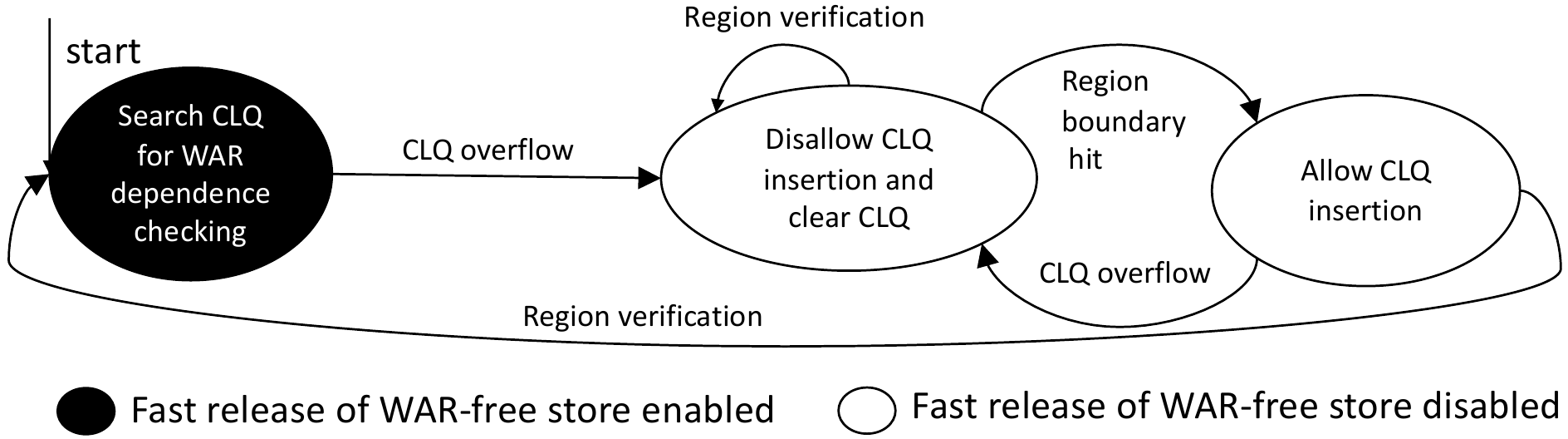}}
    \caption{Selective control for WAR-free stores' fast release}
    \label{fig:selectively_anti_dependence_checking}
\end{figure}

Once each region gets verified, \name clears only the CLQ entries that were
populated during the execution of the region. In particular, if the CLQ is
full, \name does not stall the pipeline. Whenever CLQ overflows, it instead disables the fast
release logic for WAR-free stores, i.e., load address insertions to CLQ are
blocked and it is wiped out, making the following stores go through the SB
quarantine as is. 
When a new region starts thereafter, \name resumes the CLQ insertion so that the region can leverage the fast release of its WAR-free stores unless the CLQ overflows. 
More precisely, to ensure in-order store release to L1 cache, 
\name does not enable the fast release logic until the prior region is verified with its stored released. 
Figure~\ref{fig:selectively_anti_dependence_checking} illustrates how \name
selectively controls (enables/disables) the fast release of WAR-free stores according to the CLQ
status, i.e., whether it is full or not.

\paragraph{Compact CLQ Design with Range Checking}
In general, the bigger CLQ size is, the more often the fast release logic is
enabled---leading to more WAR-free stores that can be merged to cache without
their region verification. However, we found out that the WAR dependence is rarely found in
each region. Taking this into account, we propose a range-based address checking
that can compress all the addresses of the loads executed in each region by
keeping the range of the minimum and maximum addresses during the execution.
In this way, \name only needs to allocate a single CLQ entry for each region without hurting the precision significantly.

Furthermore, such a per-region range-based CLQ entry renders the
WAR dependence checking logic faster and simpler, which would otherwise require
CAM (content-addressed memory) search for multiple entries. That is, for each regular store of a given
region, \name (1) looks up the CLQ entry corresponding to the region and (2) checks if the store
address falls into the address range of the entry.  The upshot is that \name
can significantly reduce the hardware cost for CLQ with neither the significant loss of
the precision to detect WAR-free stores nor the visible performance degradation compared
to the address matching based ideal CLQ.

\begin{figure}[h!]
    \centering
    \includegraphics[width=1.0\columnwidth]{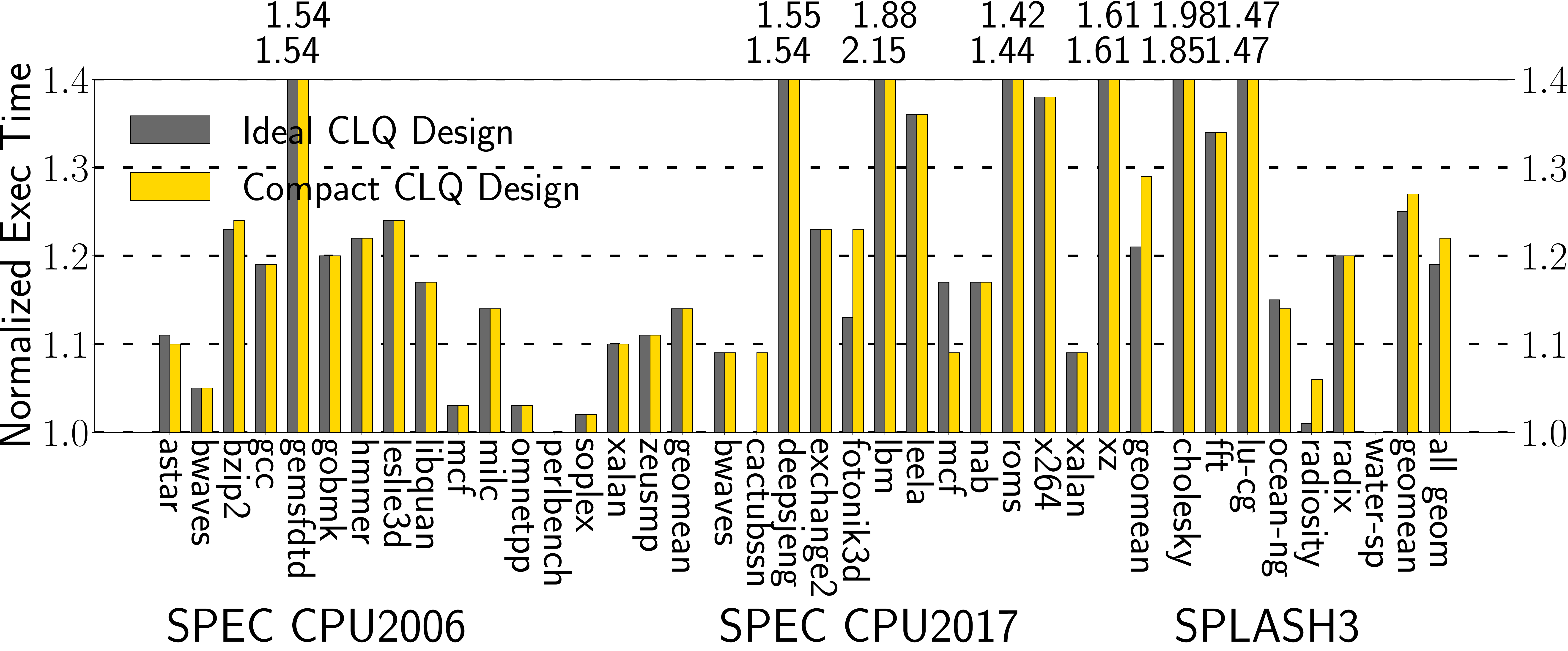}
    \caption{Run-time overhead compared to original program without
             resilience support between an ideal CLQ
             (infinite-size CLQ) and \name's compact 2-entry CLQ;
             note that we only enable WAR-free checking and hardware coloring to
             exclude the impacts of \name compiler optimizations}
    \label{fig:benefit_compact_clq}
\end{figure}

\begin{figure}[h!]
    \centering
    \includegraphics[width=1.0\columnwidth]{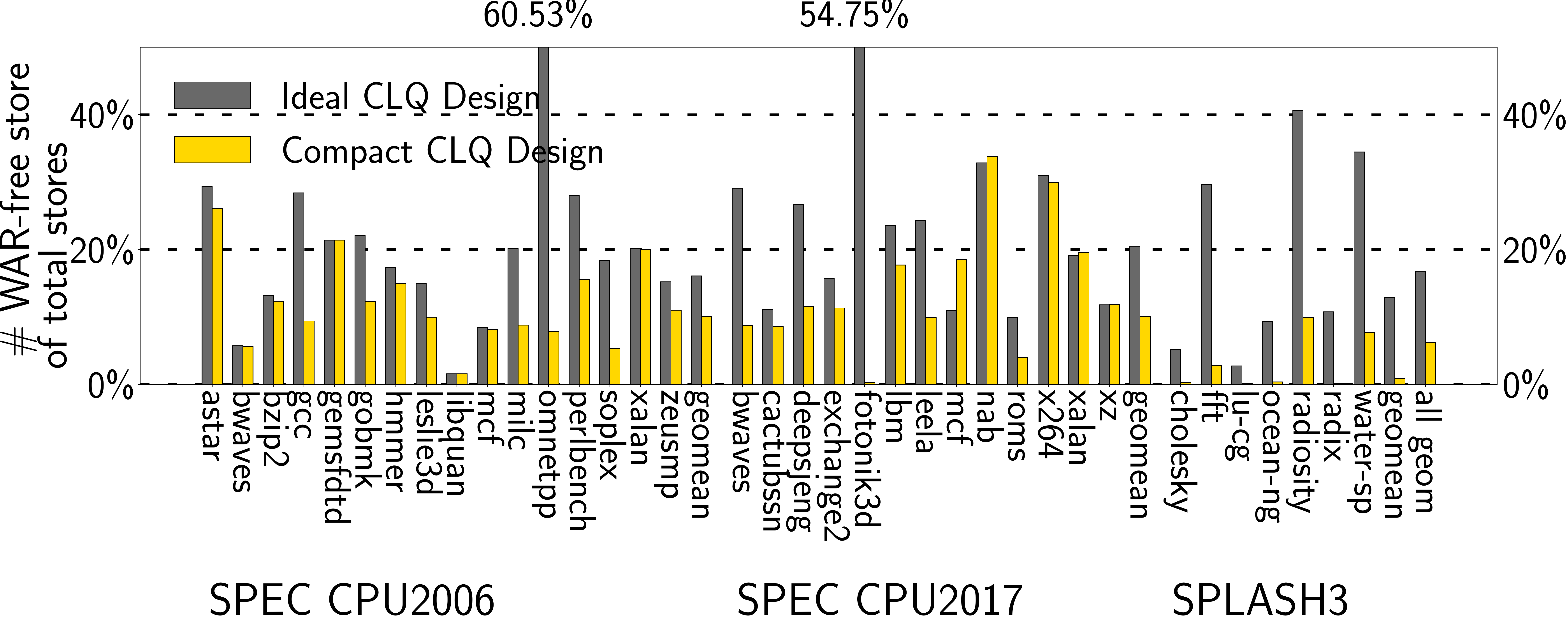}
    \caption{Ratio of detected WAR-free stores to all stores including checkpoints (higher is better)
             for the ideal basic CLQ (infinite-size) and \name's compact CLQ (2 entries)}
    \label{fig:accuracy_compact_clq}
\end{figure}

To confirm this, we compare the performance of \name's compact CLQ
against the ideal (100\%-accurate) CLQ that
performs address matching to identify WAR-free stores with an infinite
number of CLQ entries. Figure~\ref{fig:benefit_compact_clq} shows the performance
overhead of the 2 designs which is normalized to the original application
execution time that has no soft error resilience.  It turns out that \name's
compact CLQ design only incurs 3\% performance loss on average compared to the
infinite-size ideal CLQ.  As shown in
Figure~\ref{fig:accuracy_compact_clq}, that is because the infinite-size ideal
CLQ leads to 10.58\% higher detection accuracy than \name's compact CLQ.

Finally, since \name's compact CLQ has only 2 entries by default, it is
technically possible to encounter the CLQ overflow, that is handled by the
selective fast release control (Figure~\ref{fig:selectively_anti_dependence_checking}). For example, suppose that \name executes three consecutive regions; while the first region is being verified, \name can reach the end of the second region in which case
 CLQ does not have an available entry---due to the overflow---for accommodating the addresses of the last region's loads. In fact, \name's compiler ensures that each region cannot have more than
 half of the SB size so that the verification of one region can be overlapped
 with the execution of the next region.  However, since the region
 partitioning~\cite{liu2016low} is based on a path-insensitive
 analysis~\cite{muchnick1997advanced}, some regions might have even a smaller number
 of stores than the half of the SB size, e.g., regions could have only one
 store when the SB has 4 entries as with ARM Cortex A53. As will be shown in
 Figure~\ref{fig:clq_size}, the compact CLQ needs 3-4 entries 
 to prevent the overflow for all our benchmarks.

\subsubsection{Fast Release of Checkpoint Stores}
\label{sec:color}
By definition, all checkpoints are a WAR-free store in their region because
the register stored by a checkpoint is never read by its own region; rather,
the register is only used as an input to some later region. For this reason,
one might think it is ok to release checkpoint stores without the SB quarantined
for verification. However, we found it impossible because the error recovery
could fail sometimes.

\begin{wrapfigure}{R}{0.6\columnwidth}
    \centering
    \includegraphics[width=0.6\columnwidth]{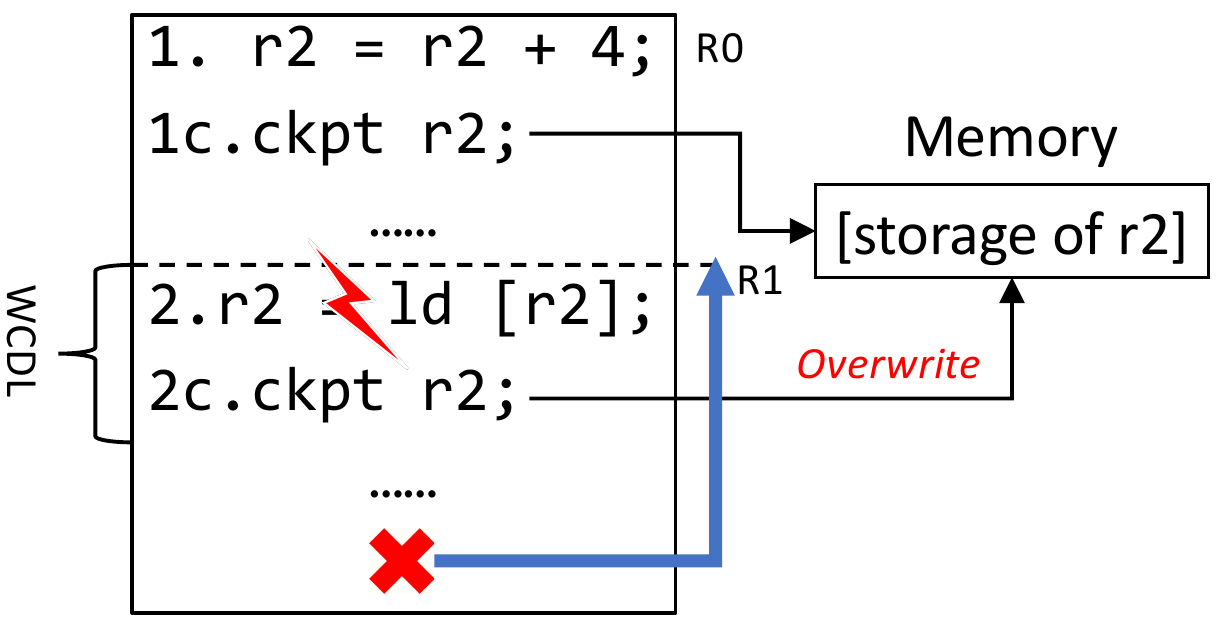}
    \caption{Problem of releasing of checkpoints without verification}
    \label{fig:hardware_coloring_motivation}
\end{wrapfigure}

Figure \ref{fig:hardware_coloring_motivation} describes such a corner case
with two regions R0 and R1 that both checkpoint the same register
$r2$.  Suppose $r2$ at line 2 is corrupted due to a soft error, and the error is
detected after the verification point of the prior region R0. That is, the next region
R1 is to be re-executed for the error recovery.  Here, if a checkpoint of $r2$
at line 2c is merged to cache without verification---though it is corrupted,
the checkpoint memory location is going to be overwritten by the corrupted value of $r2$.
Hence, the re-execution of R1 ends up restoring its input register $r2$
from the corrupted value, thereby failing to correct the error.

The crux of the problem is that the checkpoint storage (location) is
overwritten. With that in mind, we prevent the overwriting with alternative storage.
This allows \name to safely release even checkpoint stores
immediately after their commit bypassing the verification.  To achieve this,
\name leverages simple microarchitectural support called hardware coloring
that can dynamically assign a distinct memory location (i.e., color) to a
checkpoint. As such, \name prepares a coloring pool, i.e., a set of
memory locations as checkpoint storages, to manage the available colors for each register.

\paragraph{Implementation Details of Hardware Coloring}

To ensure that a distinct color is assigned to each checkpoint, \name prepares
a 4-color pool, i.e., there are 4 checkpoint memory locations for each
register, and manages 3 register maps: Available\_Colors (AC), Used\_Colors
(UC) and Verified\_Colors (VC). For a given register, AC maps it to the next
available color while UC to the color that is used
(assigned) for each region; \name maintains UC as a part of RBB entry as shown in
Figure~\ref{fig:hardware}.  Likewise, VC maps a given register to the
verified color (the checkpoint storage)---from which \name restores the input
register of the region being restarted on recovery. 

\begin{figure}[ht!]
    \centering
    \includegraphics[width=1.0\columnwidth]{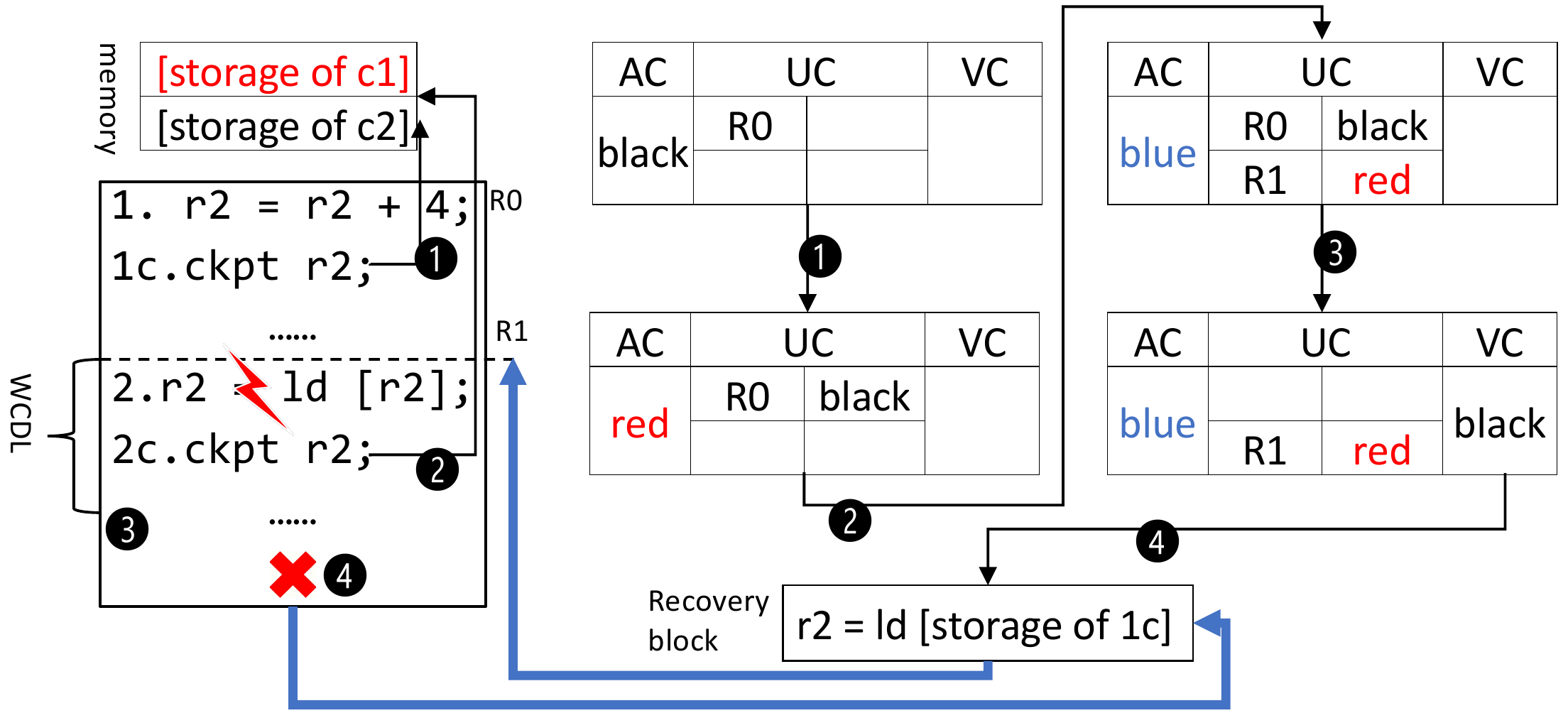}
    \caption{Fast release of a checkpoint store}
    \label{fig:hardware_coloring_example}
\end{figure}

Initially, VC is empty since there is nothing verified. When the in-order pipeline encounters a checkpoint,
\name tries to assign a color to the checkpoint by referring to AC with the
register being checkpointed. If there is an available color, it is
inserted to the UC of the region in which the checkpoint store exists;
otherwise, \name simply gives up the fast release
of the checkpoint store and falls back to the store buffer quarantine for verification.

Figure~\ref{fig:hardware_coloring_example} shows how the status of AC, UC, and
VC changes for register $r2$ being checkpointed. Here, checkpoint at line $1c$ is assigned 
\textbf{black} from AC, and thus the UC of a region R0 is updated with \textbf{black} (\ding{182} in the figure). 
Similarly another checkpoint $2c$ is assigned \change \textbf{red}\stopchange, and
the corresponding register mapping in the UC of region R1 is updated with \change \textbf{red} \stopchange (\ding{183}).

Once a region is verified, \name flushes every color of VC to AC for reclamation purpose
and updates the VC with the used colors of the verified region which are obtained by
searching the UC with the region as a key. As shown in Figure~\ref{fig:hardware_coloring_example}, once R0 is
verified at the end of WCDL after it is finished, \name updates the VC with the color(s) 
that UC holds for R0, i.e., \textbf{black} (\ding{184}).

When an error is detected in a region R1 at some point after WCDL (\ding{185}) in
the figure, \name invokes the recovery block to restore the value of the input
register $r2$ from the \textbf{black} checkpoint storage, and then jumps back
to the recovery PC, i.e., the entry of R1. 
Overall, \name's hardware cost is not significant as will be shown in Section~\ref{sec:area_power_overheads}. 


\section{Discussion}
\label{sec:discuss}
\noindent {\bf Fault Model: } We assume that both SB and RBB are hardened to be
	robust against soft errors as in prior work~\cite{liu2016low}. Besides, the 2
	entries of the committed load queue (CLQ) and the three color maps (total 6
			bits per register) are to be protected. Like prior work and commodity RAS (reliability/availability/serviceability)
	processors, caches and the address generation unit (AGU) should be hardened.
	Finally, a single parity bit is necessary for each register in case it holds store's address operand whose corruption ends up altering random memory location under \name's fast release.
	\name prevents this problem by causing any parity-detected error upon each register access to trigger its recovery process---as if it were detected by acoustic sensors.

\noindent {\bf Store Buffer Scaling: }
In general, it is challenging to enlarge a store buffer because
SB's store-to-load forwarding impacts the length of a pipeline clock tick.
SB must provide data within L1-hit time to avoid complications of scheduling loads
with variable latency, e.g., for a 16/32-entries SB in Alpha AXP processor
clocked at 3GHz, the store-to-load forwarding latency increases to 3-4
cycles\cite{sha2005scalable}. Especially for in-order cores, it is even more
challenging due to the power-hungry nature of the CAM (content-addressed memory)
search for the store-to-load forwarding. That is why commodity in-order cores
have only a few SB entries, e.g., ARM Cortex-A53 has 4 entries. In addition, 
Section \ref{sec:area_power_overheads} discusses the area and energy overheads of a large SB design
in detail.

\ignore{
\noindent {\bf Exception Handling: }
\name checks whether an exception is not the result of prior soft errors, thus
purposely postponing the handler invocation by waiting for WCDL cycles.  If
an error is detected during the WCDL time, \name triggers its recovery as is,
i.e., by re-executing from the recovery PC to the point where the
exception took place. Subsequently, the exception handler is invoked to
deal with the exception.

\noindent {\bf Multi Cores: }
Liu~\emph{et~al}~\cite{liu2016low} demonstrate that Turnstile works for multi-core
systems, treating fences and atomic operations as region boundaries.
Thus, Turnpike also works for them because it piggy-backs on
Turnstile. Although \name fast releases some stores of multithreaded program, the stores are still sequentially released---from SB---to L1 cache without violating the memory consistency model. 
The upshot is that \name works for data-race program without significant performance degradation.
}
\ignore{
It is important to note that in-order only multi-core processors
are rare. To the best of our knowledge, the most recent commodity example is
Atom N330~\cite{roberts2009arm, singh2010atom} that was released in 2008.
For this reason, this paper evaluates Turnpike with sequential benchmarks---though
there is no technical challenge for running parallel ones.}

\section{Implementation and Evaluation}\label{sec:evaluation}
\ignore{
This section attempts to answer the following questions.
\begin{itemize}
\item How do \name's optimizations reduce the overall overhead for different cycles of WCDL?
\item How much does each optimization contribute to performance improvement?
\item How accurate are the optimizations for removing unnecessary checkpoints
and releasing safe stores from the SB without the quarantine for verification?
\item What are the area and power overheads of \name?
\item How sensitive is \name to different SB/CLQ sizes?
\revision{
\item What are average region size and code size increase?
\item How do LIVM and Store-Aware Register Allocation affect instruction count? 
}
\end{itemize}
}

\subsection{Methodology}
\begin{wrapfigure}{R}{0.4\columnwidth}
    \centering
    \includegraphics[width=0.4\columnwidth]{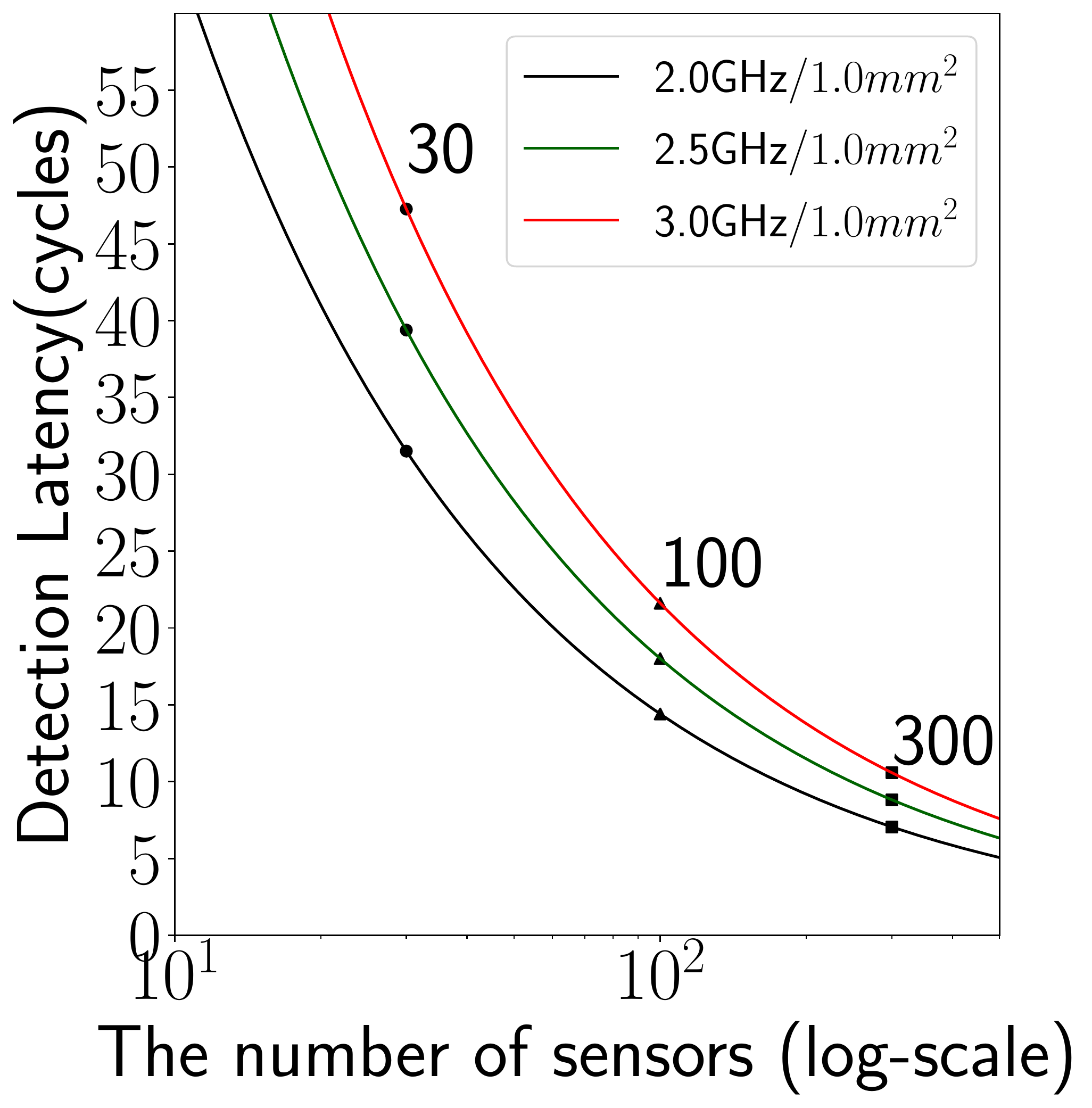}
    \caption{Detection latency across the number of deployed sensors}
    \label{fig:latency_num_sensors}
\end{wrapfigure}

\begin{figure*}[t!]
\includegraphics[width=1.0\textwidth]{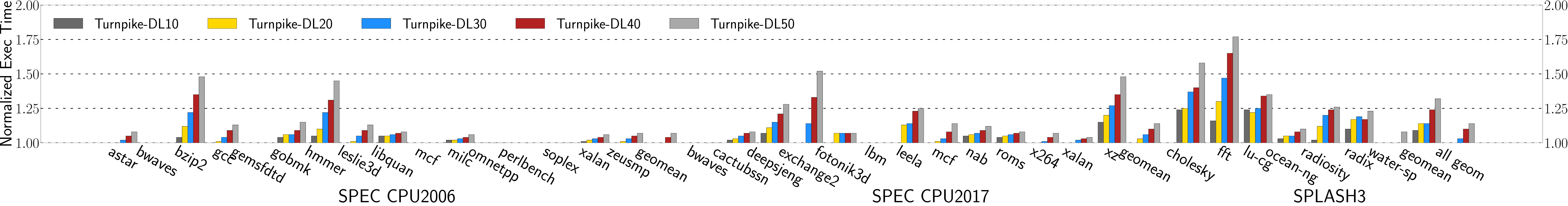}
\caption{Performance overhead of \name with varying WCDL from 10 to 50 cycles}
\label{fig:turnpike_overall_overhead_wcdl}
\end{figure*}

\begin{figure*}[t!]
\includegraphics[width=1.0\textwidth]{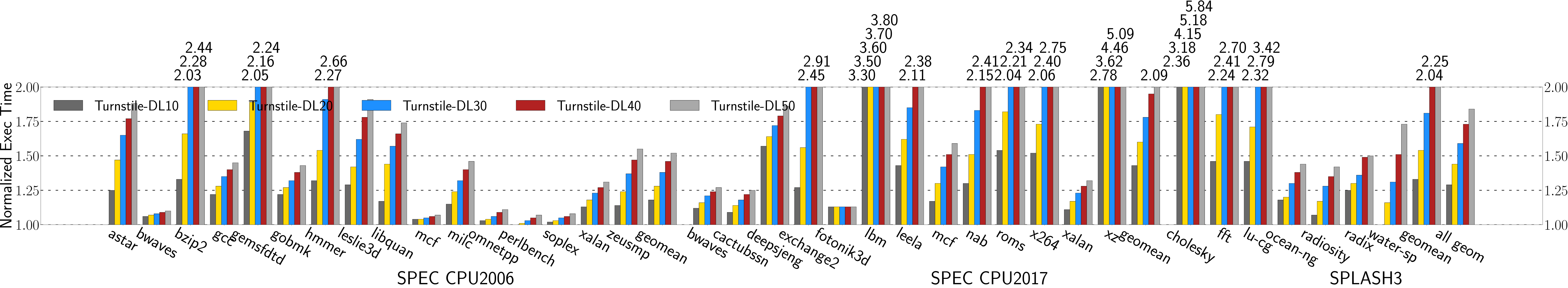}
\caption{Performance overhead of Turnstile with varying WCDL from 10 to 50 cycles}
\label{fig:turnstile_overall_overhead_wcdl}
\end{figure*}


We implemented our optimizations presented in
Section~\ref{sec:overview} with LLVM compiler~\cite{lattner2004llvm}. 
To evaluate \name's performance,
	 we used SPEC2006\cite{henning2006spec}/SPEC2017\cite{bucek2018spec} and SPLASH3\cite{splash3} compiling all benchmarks with -O3.
We conducted simulation using gem5~\cite{gem5} which is configured with
2-issue, 2.5GHz dual-core processor with 32KB/64KB 2-way set-associative L1
instruction/data caches (2 cycles hit) and a unified 128KB 16-way
set-associative L2 cache (20 cycles hit) to
model an ARM Cortex-A53 processor~\cite{A53}. The store buffer size is set to 4
as with the recent work that simulates the Cortex-A53
core~\cite{jeong2020casino}, and the default CLQ size is 2. According to prior works~
\cite{UpasaniVG12,UpasaniVG13,UpasaniVG14,UpasaniVG15}, 300-30 deployed
acoustic sensors can achieve 10-30 cycles of the worst-case detection latency
(WCDL) with the area cost of less than 1\% of die size, and therefore we set the
default WCDL to 10 cycles.

For SPEC CPU2006 and SPEC CPU2017, we synchronized the number of
simulated instructions by measuring the number of the function call
instructions which is a constant across binary versions
generated by different compiler optimizations. All benchmarks were
fast-forwarded through the number of function calls to execute at least
5 billion instructions on the original executable without soft
error resilience support, then we simulated the next 1 billion instructions
with the gem5 in-order pipelined processor model. To be more practical, we
simulated all SPEC CPU benchmarks with the reference inputs. For SPLASH3 benchmarks, 
				we simulated the entire program with full system model of gem5.
In the following, all performance results are presented as a slowdown,
i.e., the inverse of a speedup, to the baseline that has no soft error resilience support.

\subsection{Run-time Overhead with Varying WCDL}
WCDL (worst-case detection latency) is inversely proportional to the number
of sensors deployed and affected by the underlying clock frequency; the higher frequency the clock is, the longer the WCDL is.
Figure~\ref{fig:latency_num_sensors} shows these trends for 300-30 sensors
deployed on top of $1 mm^2$ core die, e.g., 10 cycles WCDL for 2.5GHz core
with 300 sensors.  Due to the process technology and the fabrication issue,
deploying all 300 sensors might not be possible under the budget of 1\% die
size overhead.  Thus, we vary WCDL from 10 cycles up to 50
cycles to cover other possible fabrication cases and evaluate the 
general trend of \name's overhead across the different WCDLs.

Figure \ref{fig:turnpike_overall_overhead_wcdl} presents the run-time overhead of
\name for 5 WCDLs: 10/20/30/40/50.
\name incurs only 0-\revision{14\%} overheads on average for the varying WCDLs from 10 to 50 cycles. In contrast, Turnstile suffers 29-84\% average overheads for the 5 WCDLs (see Figure \ref{fig:turnstile_overall_overhead_wcdl}). It is worth noting
that \name significantly outperforms Turnstile for all the benchmarks. In particular, when 10-cycle WCDL is used by default, 
	 \name's overhead is only around 1\% for most of the benchmarks, thereby delivering 0\% average overhead! 

\ignore{
\subsection{Impact of Compiler Optimizations and HW Support}
Figure \ref{fig:compiler_opts_hardware_opts} demonstrates the performance impact
of \name's compiler optimizations and hardware-supported techniques for the same series of the WCDLs.
It turns out that the resulting overhead increases sharply up to 34\% from 3\% across the WCDLs \revision{with
only compiler optimizations enabled}, while the hardware techniques show a more gradual overhead increase.
In particular, it turns out that the compiler optimizations and
the hardware techniques work in a synergistic manner. That is, their combination (i.e., \name) always
beats the compiler-only and hardware-only techniques. 
}

\vspace{-10pt}
\subsection{Impact of \name's Optimizations}\label{sec:optimizations_effect}
\begin{figure*}[t!]
\includegraphics[width=1.0\textwidth]{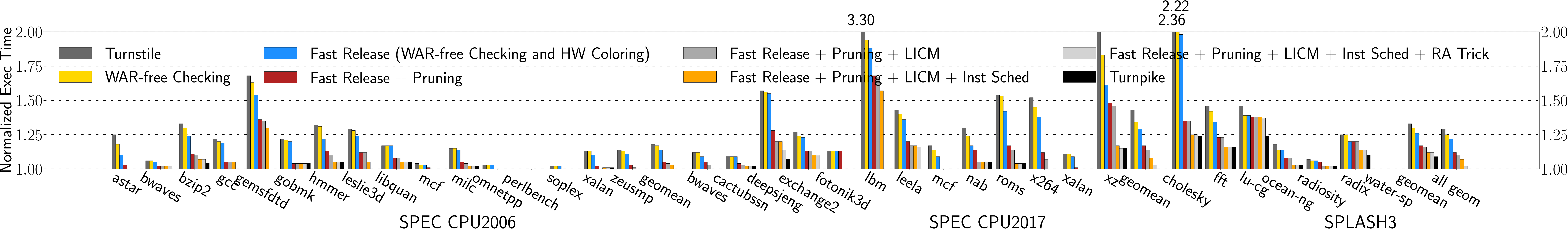}
\caption{Performance comparison between Turnstile and \name's optimizations with 10-cycle WCDL.}
\label{fig:overall_overhead}
\vspace{-10pt}
\end{figure*}

\begin{figure*}[h]
\includegraphics[width=1.0\textwidth]{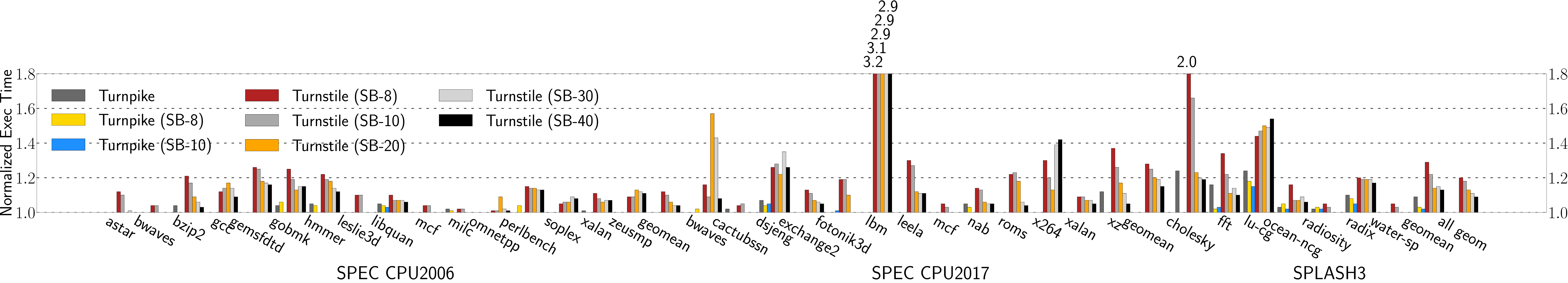}
\caption{Performance comparison of \name and Turnstile with different SB sizes
        (8, 10, 20, 30, 40) using 10-cycle WCDL}
\label{fig:sensitive_to_sb}
\vspace{-10pt}
\end{figure*}

This section presents the performance impact of \name's optimizations.
Figure \ref{fig:overall_overhead} shows the performance results of the
following 8 cases for the default 10-cycle WCDL.

\noindent{\textbf{Turnstile:}} is the state-of-the-art work, that does not use our optimizations, and incurs a 29\% overhead on average.

\noindent{\textbf{WAR-free Checking:}} uses the fast release of regular stores; it reduces the overhead to 25\%.

\noindent{\textbf{Fast Release (WAR-free checking and HW coloring):}} enables the fast release of both regular stores and checkpoint stores; this reduces the overhead to 22\%.

\noindent{\textbf{Fast Release + Pruning:}} is the combination of the above fast release and checkpoint pruning, achieving 12\% overhead; the latter removes many unnecessary checkpoints (Figure~\ref{fig:hardware_accuracy}).

\noindent{\textbf{Fast release + Pruning + LICM:}} is the combination of the fast release, checkpoint pruning, and LICM, achieving 10\% overhead. LICM particularly works well for {\tt deepsjeng}, {\tt fotonik3d}, {\tt nab}, and {\tt x264}, reducing their overhead by $>$5\%.

\noindent{\textbf{Fast release + Pruning + LICM + Inst Sched:}} is the combination of the fast release, checkpoint pruning, LICM, and instruction scheduling. The resulting average overhead is 7\%.

\noindent{\textbf{Fast release + Pruning + LICM + Inst Sched + RA Trick:}} is the combination of the fast release, checkpoint pruning, LICM, instruction scheduling, and store-aware register allocation. On average, it reduces the overhead to 2\%. Significant overhead reduction is found in 
{\tt gemsfdtd} and {\tt lbm}; as shown by Figure~\ref{fig:hardware_accuracy}, the register allocation trick eliminates the stores of the 2 benchmarks by 19\% and 17\%, respectively.

\noindent{\textbf{\name:}} uses all above optimizations along with loop induction variable merging, eliminating Turnstile's overhead completely, i.e., \name's average overhead is 0\%!
It turns out that loop induction variable merging is particularly effective for {\tt exchange2}, {\tt leela}, {\tt lu-contigous}, and {\tt radix}.

\subsection{Impact of SB Pressure Reduction Schemes}

Figure \ref{fig:hardware_accuracy} shows the detailed breakdown of all stores with
8 categories: \textbf{Pruned} corresponds to 
the checkpoint stores eliminated by the optimal checkpoint pruning while 
\textbf{LICM-eliminated} to those removed by the loop-invariant code motion.  Among the remaining checkpoint stores, \textbf{Colored} corresponds to those that can be merged to cache without the SB quarantine. Similarly, \textbf{WAR-free} 
corresponds to the regular stores that can be merged to cache without verification.
Next, \textbf{RA-eliminated} and \textbf{IndVarMerging-eliminated} correspond to
those stores that can be removed by our store-aware register allocation
and loop induction variable merging optimization respectively. Finally, \textbf{Others} represents the rest of the stores which cannot be removed or fast released by \name thus going through the verification.
As shown in the figure, the checkpoint pruning removes 21\% of all stores while LICM removes 1.4\% of them on average. Although LICM has little impact for the majority of the benchmarks, its checkpoint removal is significant for \texttt{cactubssn}, \texttt{lbm}, \texttt{cholesky}
and \texttt{radix}. Meanwhile, 1.7\% and 5\% of all stores are removed by store-aware register allocation and loop induction variable merging respectively.
Finally, 39\% of all stores can be released to cache without going through the SB quarantine, highlighting the effectiveness of \name's fast release.

\subsection{Hardware Cost Analysis}\label{sec:area_power_overheads}
\begin{table}[htb]
\footnotesize
\resizebox{\columnwidth}{!}{%
\begin{tabular}{|c|c|c|}
\hline

& Area (${\mu m}^2$)  & Dynamic access (pJ)  \\
\hline
4-entry SB (CAM)     & 621.28  & 0.43099 \\
\hline
Color maps in \name (RAM) & 36.651  & 0.02518 \\
\hline
2-entry CLQ in \name (RAM) & 24.434  & 0.01679 \\
\hline
\name in total (color maps + 2-entry CLQ)       & 61.085  & 0.04197 \\
\hline
40-entry SB (CAM)    & 3132.50 & 2.11525 \\
\hline
\name in total / 4-entry SB     & 9.8\%   & 9.7\%   \\
\hline
40-entry SB / 4-entry SB & 504\% & 497\% \\
\hline
\end{tabular}
}
\caption{Cost comparison of Turnpike and a large SB design}
\label{table:area_power_overhead}
\end{table}
\name incurs a very small hardware overhead; the 2-entry CLQ requires 16 bytes while 
the three 4-color maps (AC, UC, and VC in Section~\ref{sec:color}) need 6 bits ($3\cdot\log_2{4}$) per register---requiring 24 bytes for 32 registers as in ARM Cortex A53. In summary, \name only needs total 40 bytes for such an in-order processor.

To further evaluate the area and the power overheads of \name, we used  
CACTI \cite{shivakumar2001cacti} with 22$nm$ technology. 
Table \ref{table:area_power_overhead} highlights the area/power-efficiency of \name's compiler/architecture co-design. Compared to ARM Cortex A53's 4-entry store buffer as a baseline, \name only incurs 9.8\% area and 9.7\% energy overheads (see the second last low of the table). 
In contrast, simply increasing the store buffer size to 40 causes 504\%/497\% area/energy overheads, which is unrealistic for low-power in-order cores.

\subsection{Sensitivity Analysis}

\begin{figure}[t!]
\includegraphics[width=1.0\columnwidth]{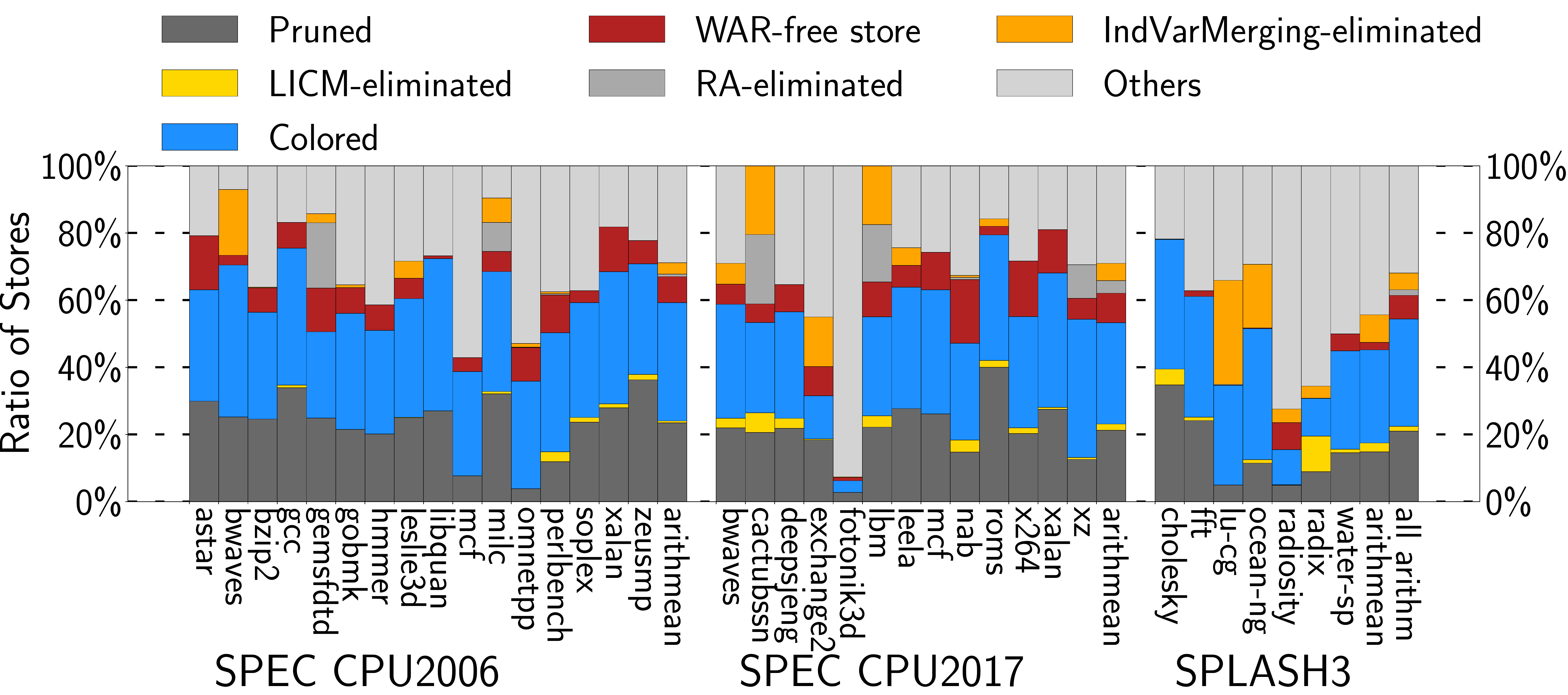}
\caption{Store breakdown; 2-entry CLQ; 10-cycle WCDL}
\label{fig:hardware_accuracy}
\end{figure}

\begin{figure}[t!]
    \centering
    \includegraphics[width=1.0\columnwidth]{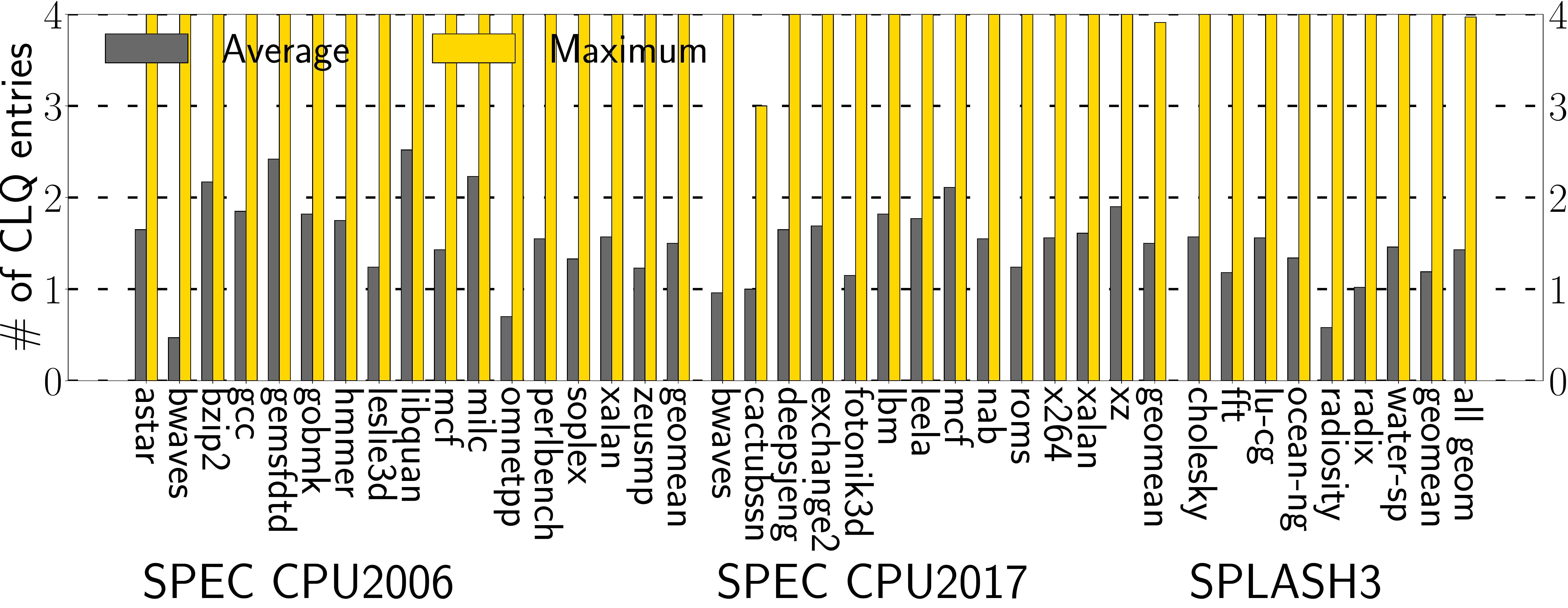}
    \caption{Dynamic CLQ entries populated; 10-cycle WCDL}
    \label{fig:clq_size}
\end{figure}

\begin{figure}[h!]
    \centering
    \includegraphics[width=1.0\columnwidth]{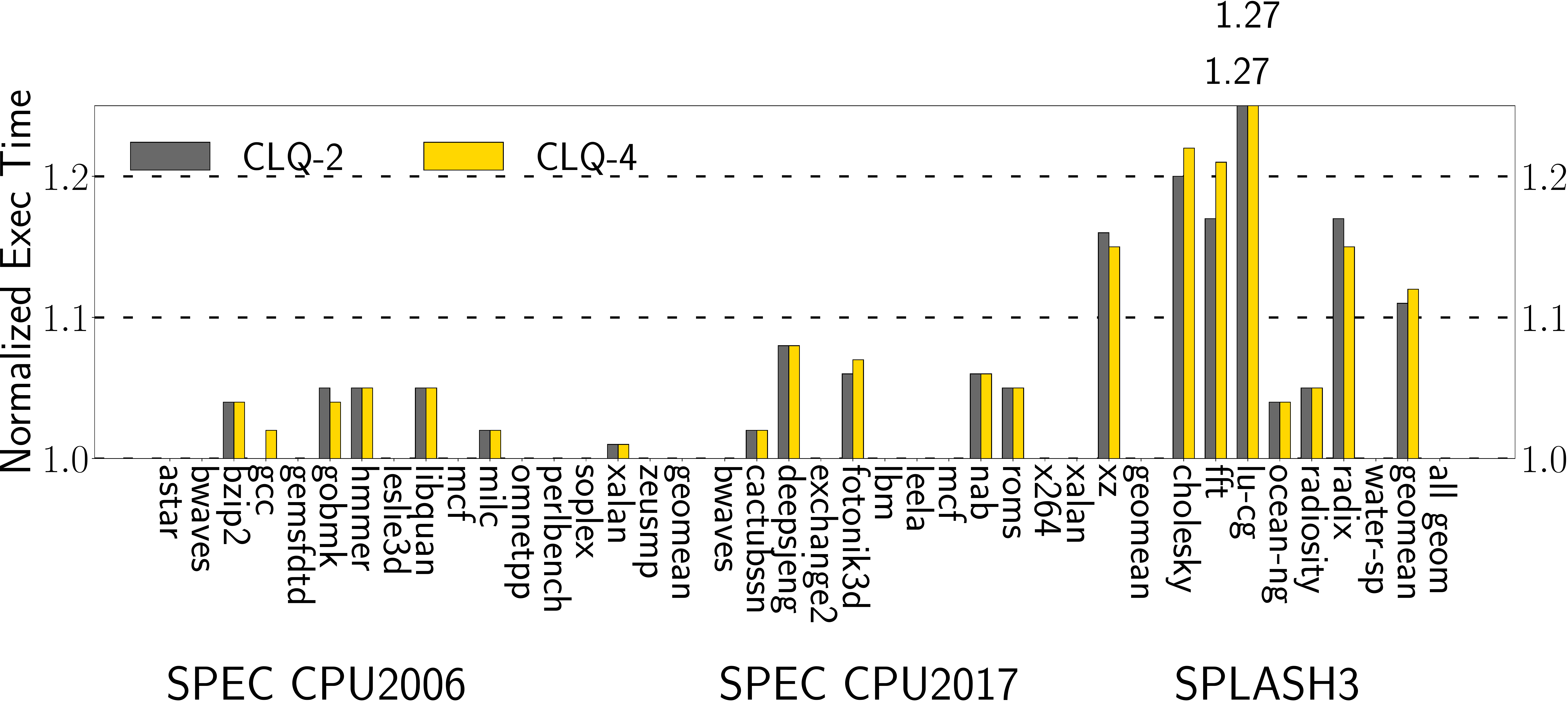}
    \caption{2-entry vs 4-entry CLQs with 10-cycle WCDL}
    \label{fig:sensitive_to_clq_size}
\end{figure}

\noindent{\bf Sensitivity to CLQ size: }
Recall that CLQ is a critical hardware structure, since its dependence checking logic is essential for the fast release of WAR-free stores. 
Figure \ref{fig:clq_size} shows the average and maximum numbers of dynamic CLQ entries populated at run time. 
\ignore{
In particular, we used the following formula to calculate the average number.
For the collected number of CLQ entries $CLQ\_entries(RB_i)$ on each region boundary $RB_i$,
average CLQ entries for an application with $N$ region boundaries can be computed as:
$$
    Average\_CLQ\_entries = \dfrac{\sum_{i=1}^{N}{CLQ\_entries(RB_i)}}{N}
$$
}
The average number of populated CLQ entries is about 1 though the maximum number goes up to 3 or 4 for some applications. Further investigation confirms that the peak number is scarcely observed. That is why \name's CLQ size is set to 2 (by default), and its performance is almost the same as that of a bigger CLQ with 4 entries as shown in Figure \ref{fig:sensitive_to_clq_size}.
The takeaway is that our compact CLQ design is not only low-cost but also high-performance.

\noindent{\bf Sensitivity to SB Size: }
It is hard to increase the size of a store buffer (SB) especially for in-order cores. Nonetheless, to highlight the performance of \name, we enlarge the store buffer of Turnstile---though it performs poorly for the 4-entry SB of ARM Cortex A53 which is \name's SB size. In addition to the default size, we tested 5 more SB sizes from 8 to 40 with 10-cycle WCDL. As shown in Figure \ref{fig:sensitive_to_sb}, for 5 SB sizes (8/10/20/30/40) with 10-cycle WCDL, Turnstile's average performance overheads are 20\%, 18\%, 13\%, 11\%, and 9\%, respectively.
Note that although Turnstile is equipped with a much larger SB, it performs significantly worse than \name. Even with the 40-entry SB that is 10x bigger than \name's SB, the average slowdown of Turnstile is 9\% whereas that of \name is 0\% (see Figure \ref{fig:overall_overhead}).
\revision{We also tested \name for bigger SB sizes. Figure \ref{fig:sensitive_to_sb} shows that the average overhead of \name is still 0\% with the SB sizes of 8 and 10 and decreases as the SB size increases.  }
\revision{
\subsection{Region Size and Code Size Analysis}\label{region_code_size}
}
\begin{figure}[h!]
  \centering
  \includegraphics[width=1.0\columnwidth]{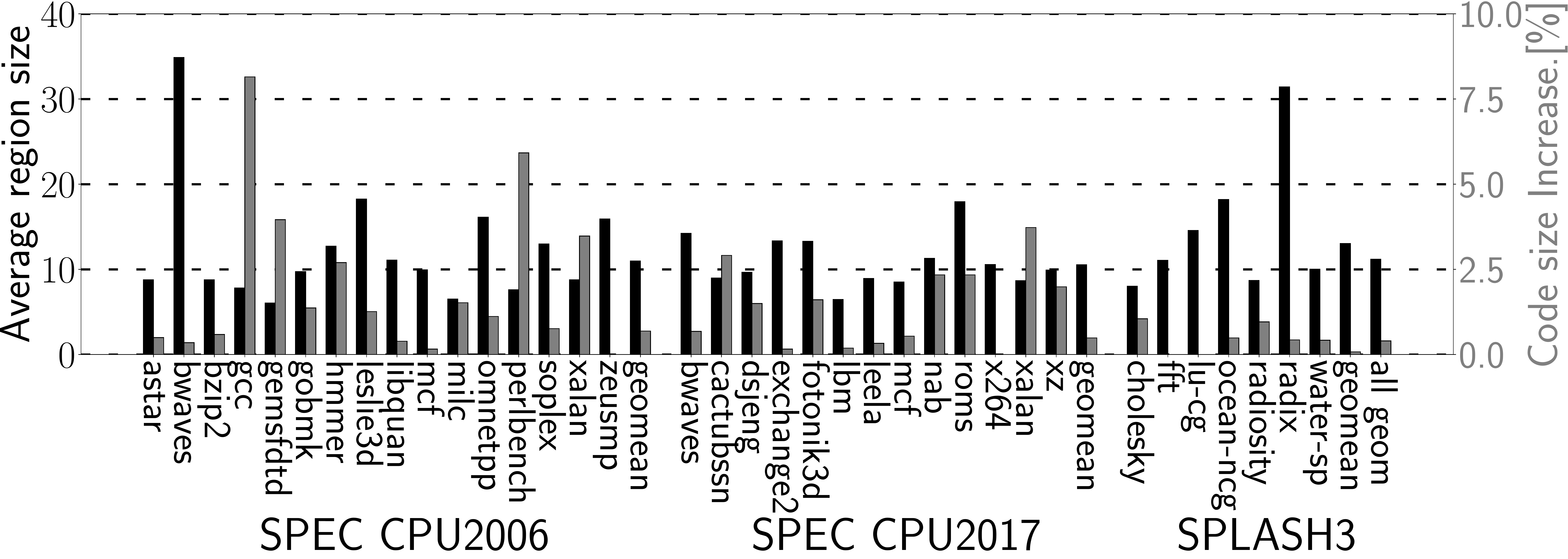}
  \caption{\revision{Region size (left) and binary overhead (right bar)
	}}
  \label{fig:region_code_size}
\end{figure}

\revision{
Figure \ref{fig:region_code_size} shows dynamic region size and binary code
size increase. On average, there are 11.2 instructions per region, and code size
increases by 0.4\% compared to the baseline. Overall, long regions lead to less
code size increase; \texttt{bwaves} has 35 instructions per region leading to 0.35\% increase, while \texttt{gcc} increases the size by 8.15\% due to many small regions (7.8 instructions per region).
}

\ignore{
\revision{
\subsection{Impact of LIVM and RA-Trick}
}

\begin{figure}[h!]
  \centering
  \includegraphics[width=1.0\columnwidth]{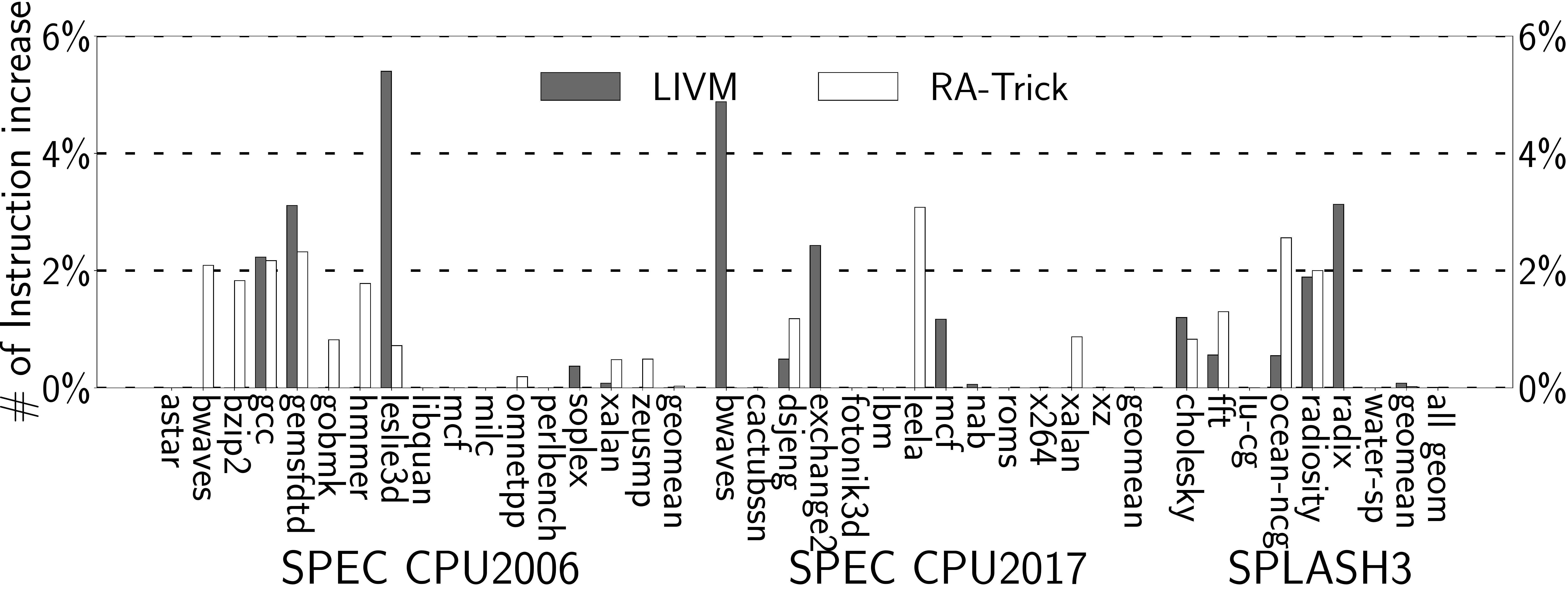}
  \caption{Dynamic instruction count increase due to LIVM and RA-Trick;lower is better}
  \label{fig:inst_count}
\end{figure}

\revision{
Figure \ref{fig:inst_count} presents that our finely tuned LIVM and Store-Aware Register
 don't increase instruction count, 0.01\% and 0.02\% respectively.
For some applications where those optimizations have observable impact (indicated as Figure
\ref{fig:hardware_accuracy}), such as \texttt{GsmFDTD}, \texttt{leslie3d}, \texttt{bwaves},
\texttt{exchange2}, \texttt{leela}, \texttt{ocean-ncg}, and \texttt{radiosity}, those
optimizations try to reduce number of stores at the cost of increasing other instructions,
which doesn't pose pressure on the small SB and thus generates negligible negative impact
on pipeline.
}
}

\section{Other Related Works}
Many prior works use redundant-computation-based detection for high error coverage.
Instruction-level duplication replicates instructions and detects errors by comparing the results of the original and replica instructions~\cite{oh2002error,reis2005swift,mitropoulou2013drift,laguna2016ipas,didehban2016nzdc,didehban2018compiler,didehban2018compiler,ma2016identification}.
In contrast, redundant multithreading simultaneously runs a redundant thread with the original thread on available cores.
Some schemes use SW techniques to realize the redundant multithreading without hardware modification~\cite{wang2007compiler,zhang2012daft,mitropoulou2016comet,so2018expert,so2019software}, while others exploit HW support to reduce the performance overhead~\cite{rotenberg1999ar,reinhardt2000transient,mukherjee2002detailed,lafrieda2007utilizing,smolens2006reunion,wang2007compiler},
Another schemes with process-level redundancy duplicate the application process to compare the outputs~\cite{shye2007using,zhang2012runtime,dobel2014can} between the parent and child processes.
Finally, non-duplication schemes detects errors by catching abnormal symptoms caused by a soft error~\cite{wang2006restore,li2008swat,hari2009low,feng2010shoestring}.

To recover from detected errors, triple module redundancy (TMR) adopts a majority voting between the 3 executions, increasing the hardware cost. The most common error recovery scheme is to use
checkpointing or logging program status (register and memory). Prior work on coarse-grained recovery requires expensive hardware support for incrementally checkpointing memory status~\cite{UpasaniVG14} or equipping the core with a large store buffer for memory logging~\cite{sorin2002safetynet}.
To reduce the checkpointing cost in a fine-grained manner, recent studies partition the program
into small idempotent regions to reduce the number of data to be checkpointed~\cite{de2013idempotent,de2012static,liu2016compiler,liu2018ido} or store-integrity regions \cite{zeng2021replaycache} for crash consistency with advent
of NVM. However, the idempotent recovery schemes still incur a significant run-time overhead due to register spilling
or checkpointing.

All prior works either impose a large hardware cost or suffer a high run-time cost.
To the best of our knowledge, Turnpike is the first, that reduces both costs effectively for in-order cores, requiring little hardware cost though it achieves almost 0\% run-time overhead.

\section{Conclusion}
This paper presents \name that achieves lightweight soft error resilience for in-order
cores with acoustic-sensor-based detection.
Using compiler optimizations and simple hardware support, \name incurs 
near-zero performance overhead.

\begin{acks}
The original article was presented in MICRO 2021 \cite{zeng2021turnpike}.
We appreciate anonymous reviewers for their constructive comments.
At Purdue, this work was supported by NSF grants 1750503 and 1814430. At SNU, this work was
supported in part by the National Research Foundation of Korea (NRF)
grants (NRF-2016M3C4A7952587 and NRF-2019M3E4A1080386) and
by the Institute for Information \& communications Technology
Promotion grant (No. 2018-0-00581, CUDA Programming Environment
for FPGA Clusters), all funded by the Ministry of Science and ICT
of Korea. ICT at SNU provided research facilities for this study.
\end{acks}
\bibliographystyle{ACM-Reference-Format}
\balance
\bibliography{references}

\end{document}